\documentclass[twocolumn]{aastex631}
\usepackage{amsmath}

\shorttitle{JWST probes star clusters at $z\simeq6$}
\shortauthors{Vanzella et al.}

\usepackage{hyperref}  
\hypersetup{colorlinks=true,linkcolor=[rgb]{1.,0.2,0.2},citecolor=[rgb]{0.1,0.4,1.},filecolor=[rgb]{0.7,0.2,0.2},urlcolor=[rgb]{0.7,0.2,0.2}}

\usepackage{color}
\definecolor{blue}{rgb}{0., 0., 1}

\newcommand {\T}{Table\,}

\newcommand{\oiidoub}{[\textrm{O}\textsc{ii}]}
\newcommand{\oiidoublam}{[\textrm{O}\textsc{ii}]\ensuremath{\lambda3727,3729}}

\newcommand{\oiiidoub}{[\textrm{O}\textsc{iii}]}
\newcommand{\oiiidoublam}{[\textrm{O}\textsc{iii}]\ensuremath{\lambda\lambda4959,5007}}
 
\newcommand{\ha}{\ifmmode {\rm H}\alpha \else H$\alpha$\fi}
\newcommand{\halam}{\ifmmode {\rm H}\alpha \lambda6563 \else H$\alpha$ $\lambda$6563 \fi}
\newcommand{\hb}{\ifmmode {\rm H}\beta \else H$\beta$\fi}
\newcommand{\hblam}{\ifmmode {\rm H}\beta \lambda4861 \else H$\beta$ $\lambda$4861 \fi}
\newcommand{\lya}{\ifmmode {\rm Ly}\alpha \else Ly$\alpha$\fi}
\newcommand{\pg}{\ifmmode {\rm P}\gamma \else Pa$\gamma$\fi}
\newcommand{\lyb}{\ifmmode {\rm Ly}\beta \else Ly$\beta$\fi}
\newcommand{\lyg}{\ifmmode {\rm Ly}\gamma \else Ly$\gamma$\fi}

\newcommand{\flyc}{\ifmmode  \mathrm{f}_\mathrm{esc}\mathrm{(LyC)} \else $\mathrm{f}_\mathrm{esc}\mathrm{(LyC)}$\fi}

\def\ergs{\ifmmode \mathrm{erg\hspace{1mm}s}^{-1} \else erg s$^{-1}$\fi}
\def\ergscm{erg s$^{-1}$ cm$^{-2}$}
\def\micron{\ifmmode \mu\mathrm{m} \else $\mu$m\fi}
\def\msun{\ifmmode \mathrm{M}_{\odot} \else M$_{\odot}$\fi}
\def\msunyr{\ifmmode \mathrm{M}_{\odot} \hspace{1mm}{\rm yr}^{-1} \else $\mathrm{M}_{\odot}$ yr$^{-1}$\fi}
\def\zsun{\ifmmode Z_{\odot} \else Z$_{\odot}$\fi}
\def\lsun{\ifmmode L_{\odot} \else L$_{\odot}$\fi}
\def\mstar{\ifmmode \mathrm{M}_{\star} \else M$_{\star}$\fi}
\newcommand{\JWST}{\textit{JWST}}
\newcommand{\HST}{\textit{HST}}

\begin{document}

\title{\JWST/NIRCam Probes Young Star Clusters in the Reionization Era Sunrise Arc}


\correspondingauthor{Eros Vanzella}
\email{eros.vanzella@inaf.it}

\author[0000-0002-5057-135X]{Eros Vanzella}
\affiliation{INAF -- OAS, Osservatorio di Astrofisica e Scienza dello Spazio di Bologna, via Gobetti 93/3, I-40129 Bologna, Italy}

\author[0000-0001-7940-1816]{Ad\'elaïde Claeyssens}
\affiliation{Department of Astronomy, Oskar Klein Centre, Stockholm University, AlbaNova University
Centre, SE-106 91 Stockholm, Sweden}

\author[0000-0003-1815-0114]{Brian Welch}
\affiliation{Department of Astronomy, University of Maryland, College Park, MD 20742, USA}
\affiliation{Observational Cosmology Lab, NASA Goddard Space Flight Center, Greenbelt, MD 20771, USA}
\affiliation{Center for Research and Exploration in Space Science and Technology, NASA/GSFC, Greenbelt, MD 20771}

\author[0000-0002-0786-7307]{Angela Adamo}
\affiliation{Department of Astronomy, Oskar Klein Centre, Stockholm University, AlbaNova University
Centre, SE-106 91 Stockholm, Sweden}

\author[0000-0001-7410-7669]{Dan Coe}
\affiliation{Space Telescope Science Institute (STScI), 3700 San Martin Drive, Baltimore, MD 21218, USA}
\affiliation{Association of Universities for Research in Astronomy (AURA), Inc. for the European Space Agency (ESA)}
\affiliation{Center for Astrophysical Sciences, Department of Physics and Astronomy, The Johns Hopkins University, 3400 N Charles St. Baltimore, MD 21218, USA}

\author[0000-0001-9065-3926]{Jose M. Diego}
\affiliation{Instituto de F\'isica de Cantabria (CSIC-UC). Avda. Los Castros s/n. 39005 Santander, Spain}

\author[0000-0003-3266-2001]{Guillaume Mahler}
\affiliation{Institute for Computational Cosmology, Durham University, South Road, Durham DH1 3LE, UK}
\affiliation{Centre for Extragalactic Astronomy, Durham University, South Road, Durham DH1 3LE, UK}

\author[0000-0002-3475-7648]{Gourav Khullar}
\affiliation{Department of Physics and Astronomy and PITT PACC, University of Pittsburgh, Pittsburgh, PA 15260, USA}

\author[0000-0002-5588-9156]{Vasily Kokorev}
\affiliation{Kapteyn Astronomical Institute, University of Groningen, PO Box 800, 9700 AV Groningen, The Netherlands}

\author[0000-0003-3484-399X]{Masamune Oguri}
\affiliation{Center for Frontier Science, Chiba University, 1-33 Yayoi-cho, Inage-ku, Chiba 263-8522, Japan}
\affiliation{Department of Physics, Graduate School of Science, Chiba University, 1-33 Yayoi-Cho, Inage-Ku, Chiba 263-8522, Japan}

\author[0000-0002-5269-6527]{Swara Ravindranath}
\affiliation{Space Telescope Science Institute (STScI), 3700 San Martin Drive, Baltimore, MD 21218, USA}

\author[0000-0001-6278-032X]{Lukas J. Furtak}
\affiliation{Physics Department, Ben-Gurion University of the Negev, P.O. Box 653, Be'er-Sheva 84105, Israel}

\author[0000-0003-4512-8705]{Tiger Yu-Yang Hsiao}
\affiliation{Center for Astrophysical Sciences, Department of Physics and Astronomy, The Johns Hopkins University, 3400 N Charles St. Baltimore, MD 21218, USA}

\author[0000-0002-5258-8761]{Abdurro'uf}
\affiliation{Center for Astrophysical Sciences, Department of Physics and Astronomy, The Johns Hopkins University, 3400 N Charles St. Baltimore, MD 21218, USA}
\affiliation{Space Telescope Science Institute (STScI), 
3700 San Martin Drive, Baltimore, MD 21218, USA}

\author[0000-0001-8057-5880]{Nir Mandelker}
\affiliation{Centre for Astrophysics and Planetary Science, Racah Institute of Physics, The Hebrew University, Jerusalem, 91904, Israel}

\author[0000-0003-2680-005X]{Gabriel Brammer}
\affiliation{Cosmic Dawn Center (DAWN), Copenhagen, Denmark}
\affiliation{Niels Bohr Institute, University of Copenhagen, Jagtvej 128, Copenhagen, Denmark}

\author[0000-0002-7908-9284]{Larry D. Bradley}
\affiliation{Space Telescope Science Institute (STScI), 3700 San Martin Drive, Baltimore, MD 21218, USA}

\author[0000-0001-5984-0395]{Maru\v{s}a Brada\v{c}}
\affiliation{Department of Mathematics and Physics, University of Ljubljana, Jadranska ulica 19, SI-1000 Ljubljana, Slovenia}
\affiliation{Department of Physics and Astronomy, University of California, Davis, 1 Shields Ave, Davis, CA 95616, USA}

\author[0000-0003-1949-7638]{Christopher J. Conselice}
\affiliation{Jodrell Bank Centre for Astrophysics, University of Manchester, Oxford Road, Manchester UK}

\author[0000-0001-8460-1564]{Pratika Dayal}
\affiliation{Kapteyn Astronomical Institute, University of Groningen, P.O. Box 800, 9700 AV Groningen, The Netherlands}

\author[0000-0001-6342-9662]{Mario Nonino}
\affiliation{INAF-Trieste Astronomical Observatory, Via Bazzoni 2, 34124, Trieste, Italy}

\author[0000-0002-8144-9285]{Felipe Andrade-Santos}
\affiliation{Department of Liberal Arts and Sciences, Berklee College of Music, 7 Haviland Street, Boston, MA 02215, USA}
\affiliation{Center for Astrophysics \text{\textbar} Harvard \& Smithsonian, 60 Garden Street, Cambridge, MA 02138, USA}

\author[0000-0001-8156-6281]{Rogier A. Windhorst} 
\affiliation{School of Earth and Space Exploration, Arizona State University,
Tempe, AZ 85287-1404, USA}

\author[0000-0003-3382-5941]{Nor Pirzkal}
\affiliation{Space Telescope Science Institute (STScI), 3700 San Martin Drive, Baltimore, MD 21218, USA}

\author[0000-0002-7559-0864]{Keren Sharon}
\affiliation{Department of Astronomy, University of
Michigan, 1085 S University Ave, Ann Arbor, MI 48109, USA}

\author[0000-0001-9336-2825]{S. E. de Mink}
\affiliation{Max Planck Institute for Astrophysics, Karl-Schwarzschild-Strasse 1, 85748 Garching, Germany}
\affiliation{Anton Pannekoek Institute for Astronomy, University of Amsterdam, Science Park 904, 1098XH Amsterdam, The Netherlands}

\author[0000-0001-7201-5066]{Seiji Fujimoto}\altaffiliation{Hubble Fellow}
\affiliation{
Department of Astronomy, The University of Texas at Austin, Austin, TX 78712, USA
}
\affiliation{Cosmic Dawn Center (DAWN), Copenhagen, Denmark}
\affiliation{Niels Bohr Institute, University of Copenhagen, Jagtvej 128, Copenhagen, Denmark}

\author[0000-0002-0350-4488]{Adi Zitrin}
\affiliation{Physics Department, Ben-Gurion University of the Negev, P.O. Box 653, Be'er-Sheva 84105, Israel}

\author[0000-0002-1722-6343]{Jan J. Eldridge}
\affiliation{Department of Physics, University of Auckland, Private Bag 92019, Auckland, New Zealand}

\author[0000-0002-5222-5717]{Colin Norman}
\affiliation{Center for Astrophysical Sciences, Department of Physics and Astronomy, The Johns Hopkins University, 
3400 N Charles St. 
Baltimore, MD 21218, USA}
\affiliation{Space Telescope Science Institute (STScI), 
3700 San Martin Drive, 
Baltimore, MD 21218, USA}

\begin{abstract}
Star cluster formation in the early universe and their contribution to reionization remains to date largely unconstrained. Here we present \JWST/NIRCam imaging
of the most highly magnified galaxy known at $z \sim 6$, the {\tt Sunrise} arc.
We identify six young massive star clusters (YMCs)
with measured radii spanning $\sim$20 pc down to $\sim$ 1 pc (corrected for lensing magnification), estimated stellar masses of $\sim 10^{6-7}$ \msun, and with ages $1-30$ Myr based on SED fitting to photometry measured in 8 filters extending to rest-frame 7000\AA.
The resulting stellar mass surface densities are higher than 1000~\msun~pc$^{-2}$ (up to a few $10^5$~\msun~pc$^{-2}$) and their inferred dynamical ages qualify the majority of these systems as gravitationally-bound stellar clusters. 
The star cluster ages map the progression of star formation along the arc, with two evolved systems ($\gtrsim 10$ Myr old) followed by very young clusters. The youngest stellar clusters ($<5$ Myr) show evidence of prominent \hb+\oiiidoub\ emission, based on photometry, with equivalent widths larger than $>1000$~\AA\ rest-frame, and are hosted in a 200 pc sized star-forming complex. Such a region dominates the
ionizing photon production, with a high efficiency $\log(\xi_{ion}[\rm Hz~erg^{-1}]) \sim 25.7$. A significant fraction of the recently formed stellar mass
of the galaxy ($10-30$ \%) occurred in these YMCs.
We speculate that such sources of ionizing radiation boost the 
ionizing photon production efficiency 
which eventually carve ionized channels that might favor the escape of Lyman continuum radiation. The survival of some of the clusters would make them the progenitors of massive and relatively metal-poor globular clusters in the local Universe.
\end{abstract}



\section{Introduction} \label{sec:intro}

Pioneering works in the pre-{\em James Webb Space Telescope} (\JWST) era have already reported on the detection of tiny star-forming regions at cosmological distances by combining {\em Hubble Space 
Telescope} (\HST) imaging and gravitational lensing 
\citep[e.g.,][]{ellis2001} and only recently recognized as globular cluster precursor (GCP) candidates  (\citealt{vanz_paving, vanz19}; see also \citealt{Welch22_clumps, elmegreen20}), along with the use of deep Very Large Telescope Multi Unit Spectroscopic Explorer (MUSE) spectroscopy performed 
on the Hubble Frontier Fields \citep[HFF; e.g.,][]{vanz_mdlf} or exploiting super-lensed arcs like the {\tt Sunburst} \citep[][]{rigby17, rivera19,vanz22_CFE,sharon2022}. 

Recently, exquisite \JWST/NIRCam and NIRISS imaging of lensed fields at unprecedented space-based angular resolution $40-140$ milli-arcsecond (mas) spanning $1-5$ $\mu m$ 
have revealed compact sources with physical properties resembling dense stellar clusters observed locally by, e.g., \cite{adamo20}.  
\begin{figure*}
\center
 \includegraphics[width=\textwidth]{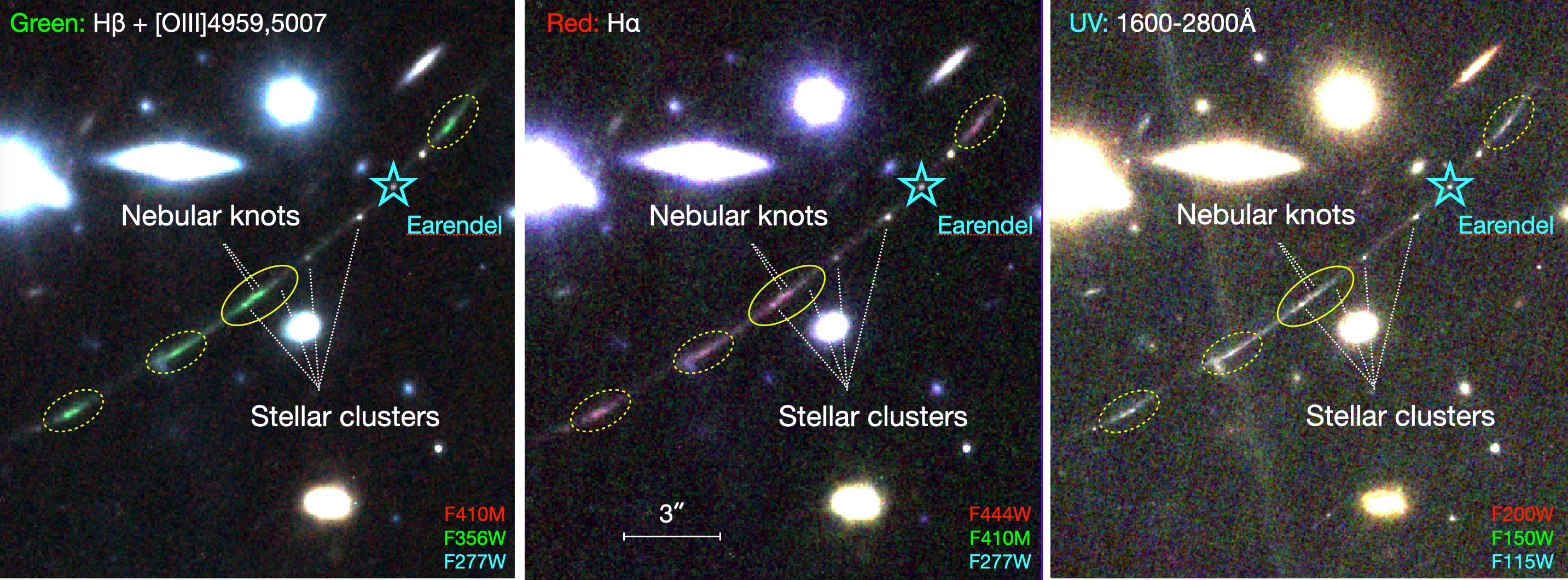}
 \caption{\JWST/NIRCAM color composites of  of the {\tt Sunrise} arc. The RGB images highlight the \hb+\oiiidoub\ (left, green channel), \ha\ (middle, red channel) and the ultraviolet continuum $\sim 1600-2800$\AA\ along the entire arc. The young massive stellar clusters and the nebular knots are indicated with dotted lines. In the bottom-right of each panel, we report the set of filters used to build the color images. Ellipses mark the four multiple images of the star-forming complex (dubbed SFC in the text). In the rightmost panel it is visible a diffraction spike which intercepts the region labeled 6c (see text and Figure~\ref{fig:summary}). Such a region is not considered in this work.}
 \label{fig:pano}
\end{figure*}
In particular, such instrumental advances led to the discovery of $z\simeq 1.37$ star clusters
with evolved ages of $0.1-4$~Gyr \citep[][]{mowla22, claeyssens22}
in a  galaxy dubbed the ``Sparkler'' which is lensed by cluster SMACS~0723, 
observed during the \JWST\ Early Release Observations \citep[ERO;][]{Pontoppidan2022}.
Relatively young proto-globular clusters at $z\simeq 4$ with estimated ages $<30$ Myr have also been identified in highly magnified galaxies by the Abell~2744 galaxy cluster \citep[][]{vanz_glass22} observed during the ERS-GLASS campaign (JWST Program 1342, PI.~T.~Treu; \citealt{TreuGlass22}). 

Such early results demonstrate that whenever the angular resolution increases, either 
by technological advancements and/or by means of gravitational lensing, the hierarchical organization of star formation promptly emerges \citep[e.g.,][]{Mestric22,johnson17} showing its fundamental units: the stellar clusters. Such young stellar clusters significantly alter the properties of the interstellar medium (ISM) of the hosting galaxy during their formation phase by injecting radiative and mechanical feedback 
into the ISM 
\citep[e.g.,][]{sirressi22_HARO11, bik18, heckman11}. These are also key processes, which might efficiently carve ionized channels and favor the escape of ionizing photons \citep[e.g.,][]{rivera19, vanz22_CFE, james16, micheva17}. 

In this work, we report on \JWST/NIRCam observations of the {\tt Sunrise} arc,
the most highly magnified galaxy known at $z \sim 6$ \citep{Salmon2020,Welch22_clumps}
and host to the lensed star {\it Earendel} \citep{Welch22_nature,Welch22_earendel}.
\cite{Welch22_clumps} analyzed the star-forming regions of the galaxy
based on \HST\ imaging.
Here we use \JWST\ to further characterize the 
same regions 
which $-$ despite the very large lensing amplification $-$ appear rather nucleated and/or barely resolved, making them some of the best candidates for globular cluster precursors (GCPs).

Throughout this paper we assume a flat cosmology with $\Omega_{M}$= 0.3,
$\Omega_{\Lambda}$= 0.7 and $H_{0} = 70\,{\rm km}\,{\rm s}^{-1}\,{\rm Mpc}^{-1}$. 
For this model, $1''=6.50$ kpc at the redshift of the cluster ($z=0.566$) and $1''=5.61$ kpc for a source at the redshift of the Sunrise arc ($z=6.2$). 
All magnitudes are given in the AB system \citep{Oke_1983}:
$m_{\rm AB} = 23.9 - 2.5 \log(f_\nu / \mu{\rm Jy})$.
In this work we use the terms star-forming {\it clumps} or {\it knots} as synonyms when referring to a sub-component of an existing galaxy, keeping in mind that the star-forming knot operational definition often applies to a source that appear nucleated or compact. With the term {\it stellar clusters} we refer to 
the usual definition of local gravitationally-bound young (massive) star clusters \citep[][]{gieles11} having a large dynamical age, high stellar mass surface density $\Sigma_M \gtrsim 1000$~M$_{\odot}$~pc$^{-2}$  and radii $\lesssim 30$ pc (see Sect.~\ref{starclusters}). Depending on the available angular resolution, a clump or knot can be a star-forming complex or a single stellar cluster, as discussed in this work.

All data analyzed in this paper are publicly available. 
We provide data products and analysis notebooks at our website.\footnote{\url{http://cosmic-spring.github.io}}

\section{\JWST\ and \HST\ imaging} \label{sec:data}

The massive galaxy cluster WHL~J013719.8–082841 ($z = 0.566$; \citealt{Wen2012,Wen2015})
was observed by the Reionization Lensing Cluster Survey \citep[RELICS,
\HST\ GO program 14096;][]{Coe_2019}
and two follow-up \HST\ GO programs 15842 and 16668 (PI Coe).
These data have recently been reprocessed and included in the CHarGE archive \citep{kokorev22}.

\JWST/NIRCam observations were acquired on July 30, 2022 
\citep{Welch22_earendel,Bradley2022highz}. 
Briefly, the cluster was observed in eight filters covering the spectral range from $0.8\mu m$ to $5\mu m$ (F090W, F115W, F150W, F200W, F356W, F410M, F444W) and with an integration time of 2104 seconds each. The superb angular resolution of \JWST/NIRcam provided PSFs with FWHM values ranging from 0\farcs04 to 0\farcs14, from 1~$\mu m$ to 5~$\mu m$, respectively. Such values were verified on a selected sample of non-saturated or uncontaminated stars identified in the same field. 

We use the \textsc{grizli} pipeline\ \citep{Grizli}
to process all \HST\ and \JWST\ individual calibrated exposures
(\texttt{FLT} and Level 2b \texttt{CAL} images, respectively).
We co-add all exposures in each filter
and align them to a common $0\farcs04$ pixel$^{-1}$ grid
with coordinates registered to the GAIA DR3 catalogs \citep{Gaia_EDR3}.
We drizzle the NIRCam short-wavelength images 
(F090W, F115W, F150W, F200W)
to higher resolution, $0\farcs02$ pixel$^{-1}$.
The \textsc{grizli} v4 data products are publicly available.\footnote{\url{https://s3.amazonaws.com/grizli-v2/JwstMosaics/v4/index.html}}
In our analysis, we use the NIRCam calibrations \texttt{jwst\_0995.pmap}, 
based on analysis of NIRCam CAL program data
and made operational on October 6, 2022.

\subsection{A super-magnified arc at z=6.0 and the emergence of tiny star-forming regions}

Thanks to the extended wavelength coverage produced by the NIRcam data we have refined the redshift estimate of the {\tt Sunrise} arc. Appendix~\ref{app_photoz} presents the
SED fitting of the host multiband photometry (see Section~\ref{app_isophot}) which results in a redshift probability distribution narrowly peaked at $z = 6.0 \pm 0.2$ (95\% CL),
with no significant likelihood at other redshifts.
This is fully consistent with previous estimates based on \HST\ imaging \citep{Salmon2020,Welch22_clumps}. In the following, we adopt the fiducial value of $z = 6.0$ as the current best solution for the {\tt Sunrise} arc (upcoming NIRSpec spectroscopy will provide definitive estimates). The results presented in this work remain the same within the uncertainties of the redshift 
estimate.

Figure~\ref{fig:pano} shows the \JWST/NIRCam color images of the {\tt Sunrise} arc along with the identified star-forming regions
as well as the star (or star system) {\it Earendel}.
The arc extends $\sim 17''$ along the tangential direction and includes multiple images of a set of tiny knots.
The total magnification is $\mu \sim 300$ based on \HST\ derived lens models \citep{Welch22_clumps}.
After minor updates to the lens models based on \JWST\ imaging as described in Appendix~\ref{app_magnif}, we find large tangential (linear) magnifications of $(30-70)\times$ for individual knots.
Such a tangential stretch corresponds to $2-4$ parsec per pixel (adopting 20 mas/pix) at $z\simeq 6.0$. The presence of barely resolved and unresolved sources despite the very large magnification implies that their intrinsic sizes are very small \citep[as observed in other systems; e.g.,][]{johnson17, rigby17, vanz19, bouwens21}. 
Small effective radii of the sources were already estimated based on \HST\ imaging
with upper limits as small as 3 pc for unresolved sources \citep{Welch22_clumps}.  
The unmatched angular resolution ($< 80$ mas at $\lambda < 2~\mu m$) and the unique wavelength coverage (up to $5~\mu m$) provided by \JWST/NIRCam allow us to 
spatially resolve all of the previously identified clumps and measure radii as small as $\sim$1 pc.

\begin{figure*}
\center
 \includegraphics[width=\textwidth]{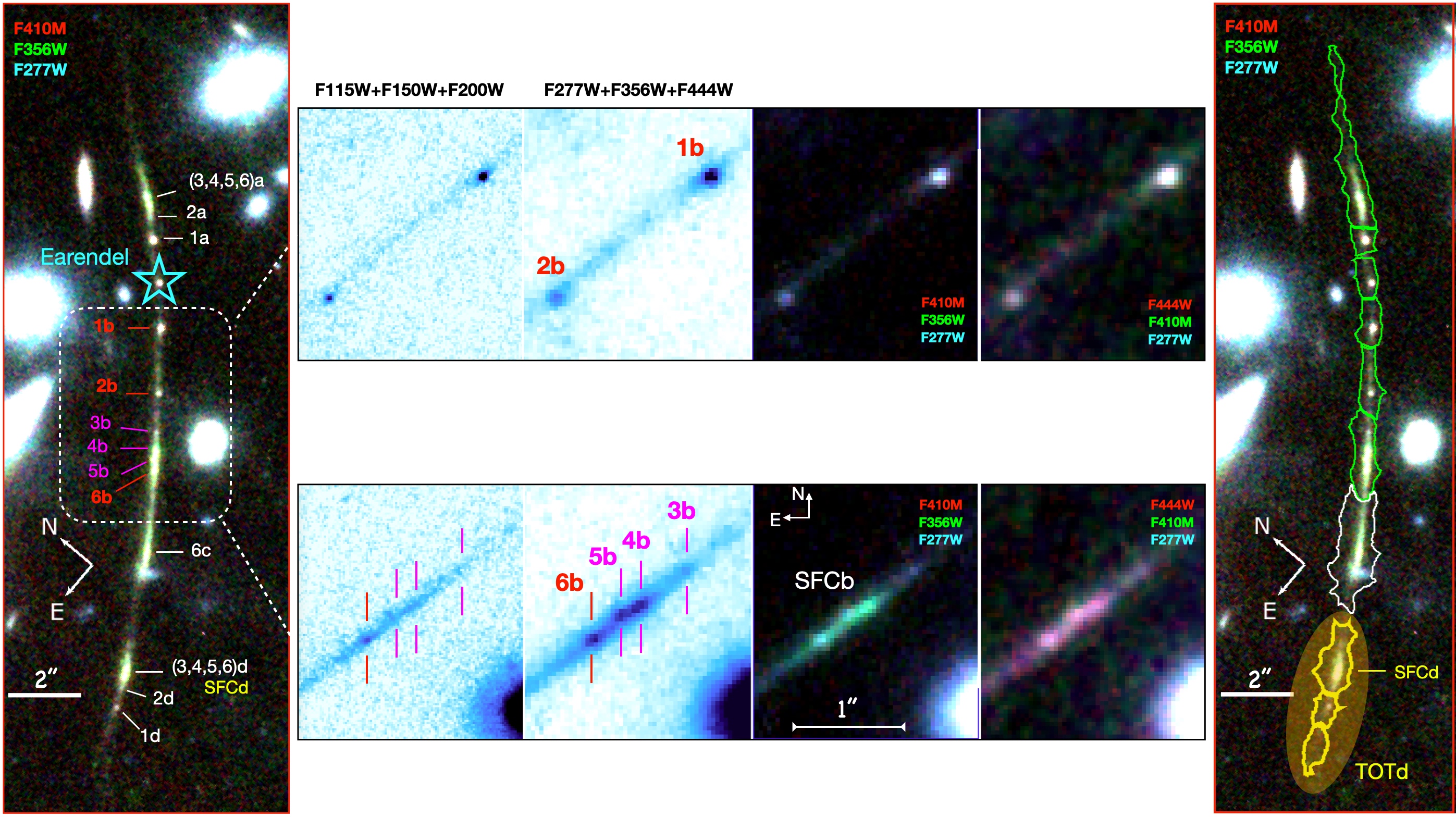}
 \caption{Summary of the detected clumps. {\it Left panel}: the {\tt Sunrise} arc is shown in the color composite \JWST/NIRCAM image in the filters F277W (blue), F356W (green) and F410M (red), with labels indicating the clumps discussed in this work. The group ``b'' represents the most magnified ones, whereas the other multiple images are indicated with white labels. Within the group ``b'' (dotted square), the red labels mark the knots identified in the ultraviolet continuum (F115W, F150W and F200W), whereas the magenta ones mark the nebular knots not detected in the ultraviolet, but which emerge where the optical nebular lines are expected to lie (F277W, F356W and F444W, see text for more details). 
 {\it Middle panels}: the thumbnails of stacked \JWST/NIRCAM bands (F115W+F150W+F200W) and (F277W+F356W+F444W) are shown. The same colored labels shown in the left panel are reported on the middle panels. The color-composite images highlight the absence/presence of optical nebular emission lines (see also Figure~\ref{fig:pano}).{\it Right panel}: the map of segments discussed in Sect.~\ref{app_isophot} are shown on the same color image shown on the left. Regarding the extended regions, we make use in this work of the segment dubbed SFCd, the combination of all yellow segments, TOTd (marked with a transparent yellow ellipse, and the combination of all segments except the white one for the estimation of the photometric redshift (Appendix~\ref{app_photoz}).}
 \label{fig:summary}
\end{figure*}

Six relatively compact knots have been visually identified after inspecting all the \JWST/NIRCAM bands. Figure~\ref{fig:summary} shows the multiple images of the {\it Sunrise} (labeled as ``a'', ``b'', ``c'' and ``d'') with the associated knots ranging from 1 to 6. In particular, objects 1b, 2b and 6b appear in all bands, from the  rest-frame ultraviolet stellar continuum (1600 -- 2800~\AA) to the optical 7000~\AA, see Figure~\ref{fig:pano}. 
In the same figure it is also visible a diffraction spike which intercepts the region labeled 6c. While such a spike does not significantly affect the photometry in that region, 6c is not
used in the rest of the work. These clumps are also clearly detected in the HST data. However, knots 3b, 4b, and 5b emerge only in the redder NIRCam bands, F277W, F356W and F444W. In the low-resolution HST data, object 3b is undetected, while 4b and 5b appear as a single clump.  As discussed in the next section, we refer to these three 
as {\it nebular} knots, as we find the photometry in the aforementioned bands could be boosted by nebular \oiidoublam, \hb+\oiiidoublam\ and \ha\ emission, respectively. These nebular knots do not show a clear compact emission in the rest-frame ultraviolet-stacked image (F115W, F150W and F200W, see Figure~\ref{fig:summary}).

The identification of high-z sources in the lensed fields depends on several effects (sizes, magnitudes, magnification, etc.) and requires dedicated simulations and/or a forward modeling approach \citep[e.g.,][]{bergamini_2022_J0416, Plazas_2019, bouwens2022_sizes}. 
We briefly address here the completeness in detecting compact sources ($<10$ pc) in the {\tt Sunrise} arc, while we postpone a full analysis to the near future when upcoming MUSE and NIRSpec spectroscopy will enable us to derive a more robust lensing model for this galaxy cluster. 
Stellar clusters with sizes smaller than 10~pc appear as (or nearly as) point-like sources in the lens plane, also in the very magnified regime studied here.  The $2\sigma$ limits of \JWST/NIRCam imaging for point-like sources in the F150W (ultraviolet rest-frame) and F410M (optical rest-frame) bands are 29.5 and 29.6 mag, respectively \citep[][]{Bradley2022highz}, which correspond to $>33.2$ and $>33.3$ adopting $\mu_{tot} > 30$ (the expected lower value along the most magnified portion of the arc, see Appendix~\ref{app_magnif}). The absolute magnitudes associated to these limits are $>-13.5$ mag and $>-13.6$ mag (they become $\sim 0.5$ magnitudes fainter if $\mu_{tot} = 50$). Such luminosities correspond to different stellar masses of the star clusters depending on the assumed Initial Mass Function (IMF), age, metallicity and dust extinction (in the ultraviolet). For simplicity, we consider the Starburst99 models \citep[][]{leitherer14}, instantaneous burst which is suitable for such tiny sources (see below), Salpeter IMF, and metallicity $Z=0.008$, where $Z$ is the mass fraction of all elements heavier than helium (though the results do not depend significantly on Z). The above ultraviolet absolute magnitude limit corresponds to a $10^{6}$~\msun\ star cluster at 40~Myr, or $0.2\times 10^{6}$~\msun\ at 10 Myr.
Given the depth of the NIRCam imaging, the geometrical configuration and the high magnification regime of the {\tt Sunrise} arc, we are likely recovering all young star clusters with stellar masses $>0.5\times 10^{6}$~\msun\ at ages younger than 10 Myr and more massive than $10^{6}$~\msun\ up to ages of 40 Myr.

\section{Physical properties of stellar cluster candidates}

\subsection{Photometry and size estimates}
\label{app_phot_knots}
Fluxes and sizes of each star cluster candidate/knot are obtained by adopting the methodology presented in
\citet[][]{claeyssens22}, following \cite{messa2022}.

\begin{figure*}
\center
 \includegraphics[width=\textwidth]{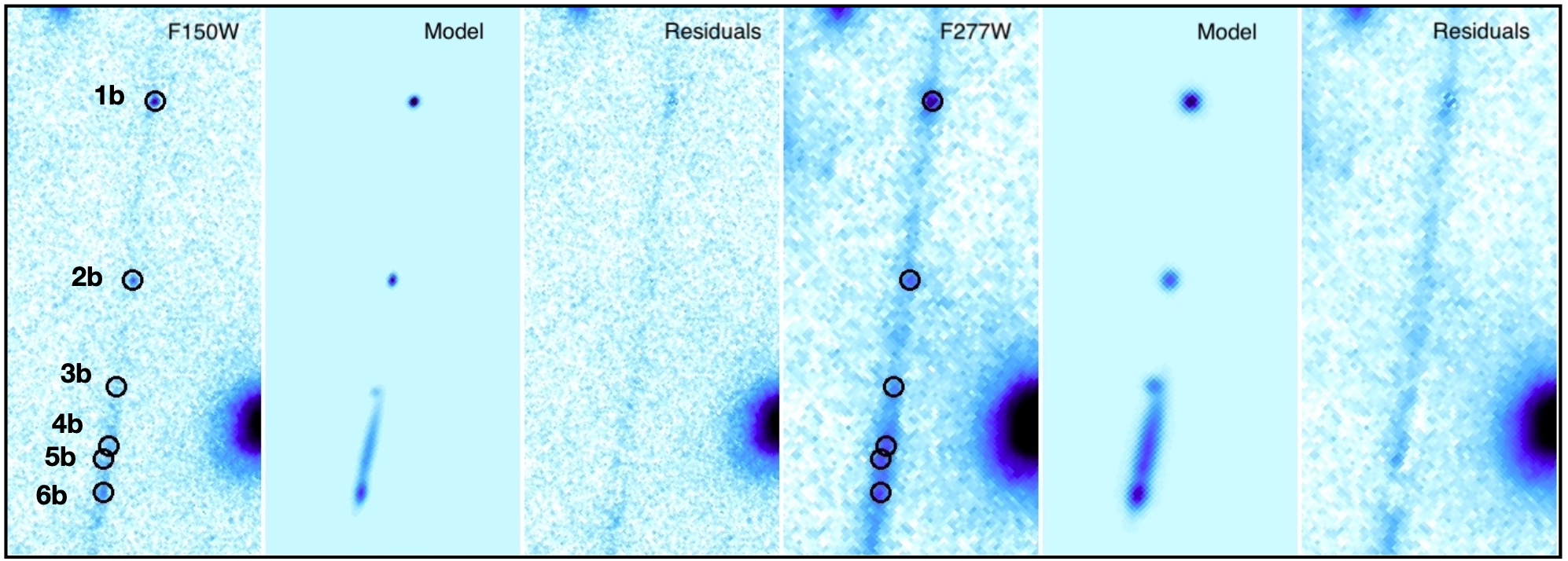}
 \caption{The resulting fitted shapes (models) and the residuals are shown for the knots of the group ``b'' reported in Table~\ref{tab:summary}. The sizes of 1b, 2b and 6b are derived from image F150W, while the nebular knots 3b, 4b and 5b from image F277W.}
 \label{fig:residuals}
\end{figure*}

Differently to clump photometric analyses based on fixed aperture photometry \citep[e.g.,][]{mowla22}, we account for the shape of the clump while measuring fluxes. We simultaneously fit for both quantities with a grid of models that are the resulting convolution of a 2D Gaussian profile\footnote{The 2D Gaussian profile is parameterized by the clump position ($\rm x_0$ and $\rm y_0$), the minor axes standard deviation ($x_{\rm std}$), the ellipticity ($\epsilon$), the positional angle ($\theta$, describing the orientation of the ellipses), and the flux ($f$).} with the measured PSF model in the image plane. The PSF is built in each \JWST\ filter by selecting and stacking observed stars in the frame. A local background gradient (i.e. emission from the galaxy light) is also fitted and removed. Since not all the clusters/knots are visible in the FUV, we use the NIRCam F150W image to determine the intrinsic shape ($x_{\rm std}$, $\epsilon$, $\theta$) of the star cluster candidates 1b, 2b, and 6b using a $9\times9$ pixel box centered on the source. For the nebular knots 3b, 4b, and 5b, we use the NIRCam F277W image as the reference filter to measure their sizes. Due to their proximity, 4b and 5b are fitted simultaneously. 
Figure~\ref{fig:residuals} shows the residuals after subtracting the models from the NIRCam images. 
Once the intrinsic shape of the object is determined we derive the fluxes in the other filters by fixing the 2D-Gaussian shape ($x_{\rm std}$, $\epsilon$, $\theta$ values), convolving it with the filter PSF, and letting the center of the Gaussian, the normalization (i.e. the integrated flux) and the local background as a free parameter. We fit the shape of the source and the underlying background within the $9\times9$ pixel box. The total flux of the source is obtained by extrapolating to infinite the best-fitted 2D Gaussian shape.
The final photometric errors include in quadrature the Poissonian error as well as the uncertainties in the measured parameters of the shape (for more details, see \citealt{claeyssens22}).
The latter error is significantly larger in the nebular knots (see Figure~\ref{fig:SEDs}), which are not well detected in the continuum bands.

The intrinsic effective radius R$_{\rm eff}$ of each source is derived in the reference frame (F150W for 1b, 2b, 6b; and F277W for 3b, 4b, 5b) by converting the x and y standard deviations of the 2D-Gaussian into a circular radius, applying the multiplicative factor to get the full-width half maximum (FWHM), and then determining R$_{\rm eff}$ as half of the FWHM. These values are reported in parsec and observed angular scale in Table~\ref{tab:summary} after distance and tangential magnification have been taken into account. In particular, the inferred angular size in the lens plane is divided by $\mu_{tang}$ and converted into parsec.
Magnitudes of each clump are reported in Table~\ref{tab:photometry}.

\begin{deluxetable*}{ccccccccccc}
\tablenum{1}
\tablecaption{\JWST/NIRCam extracted photometry of each star-forming knot and/or cluster candidate. \label{tab:photometry}}
\tablewidth{0pt}
\tablehead{
\colhead{ID} & 
\colhead{RA} & 
\colhead{DEC} & 
\colhead{F090W} & 
\colhead{F115W} & 
\colhead{F150W} & 
\colhead{F200W} & 
\colhead{F277W} & 
\colhead{F356W} & 
\colhead{F410M} & 
\colhead{F444W} \\
\colhead{} 
& \colhead{01:37} 
& \colhead{-08:27} 
& \colhead{} 
& \colhead{} 
& \colhead{}
& \colhead{} 
& \colhead{} 
& \colhead{} 
& \colhead{}  
& \colhead{}
}
\decimalcolnumbers
\startdata
1b & 23.313 & 53.13 & 27.13(0.12) & 26.85(0.11) & 26.93(0.11) & 27.03(0.1) & 26.7(0.09) & 26.59(0.09) & 26.61(0.09) & 26.57(0.09) \\
2b & 23.413 & 54.29 & 28.17(0.13) & 27.77(0.14) & 27.66(0.14) & 27.9(0.12) & 27.81(0.11) & 27.61(0.11) & 27.59(0.12) & 27.41(0.11) \\
3b & 23.474 & 54.97 & 29.2(0.26) & 29.14(0.49) & 29.42(0.47) & 29.18(0.36) & 28.25(0.24) & 28.08(0.21) & 28.74(0.35) & 28.12(0.2) \\
4b & 23.504 & 55.30 & 26.26(0.2) & 26.92(0.21) & 26.16(0.21) & 26.12(0.2) & 25.74(0.26) & 24.8(0.21) & 26.09(0.31) & 25.48(0.26) \\
5b & 23.515 & 55.40 & 29.3(--) & 29.4(--) & 29.5(--) & 29.7(--) & 28.61(0.7) & 29.12(2.28) & 28.49(0.56) & 28.15(0.67) \\
6b & 23.533 & 55.63 & 28.17(0.38) & 27.24(0.19) & 27.3(0.23) & 27.4(0.22) & 26.99(0.17) & 26.4(0.18) & 27.17(0.15) & 26.78(0.16) \\
\hline
\hline
\enddata
\tablecomments{Magnitudes and 1-sigma errors (in brackets) are reported for the NIRCam filters. Errors denoted (--) indicate lower limits at two sigma \citep[][]{Bradley2022highz}. The method to determine these values is described in Section \ref{app_phot_knots}. 
}
\end{deluxetable*}

\subsection{SED fitting analysis of star cluster candidates}

\begin{figure*}
\center
 \includegraphics[width=\textwidth]{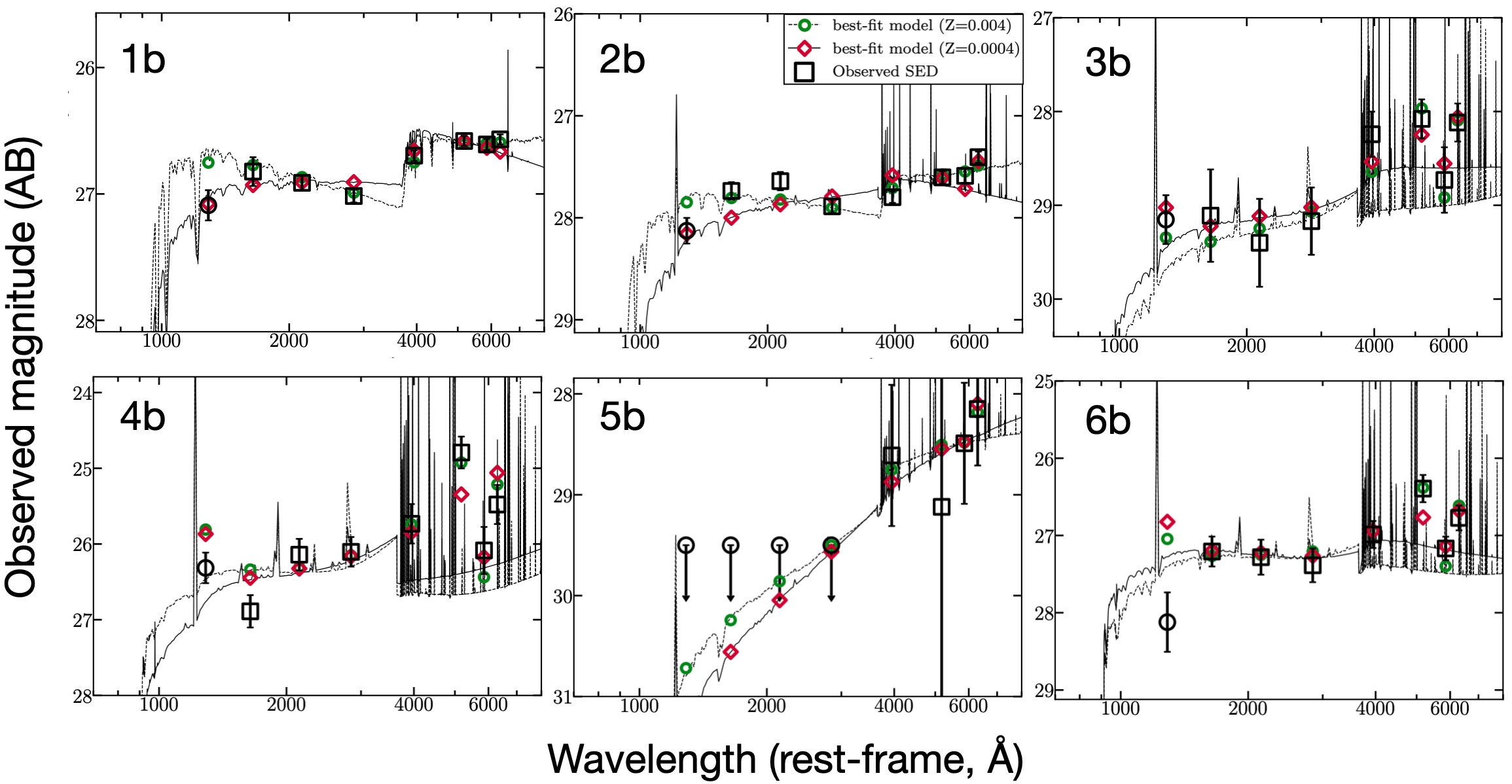}
 \caption{The SED analysis of the compact regions identified as stellar clusters and star-forming knots in the {\tt Sunrise} galaxy. Sources 1b and 2b are relatively evolved dense star clusters and do not show evident rest-frame optical emission lines, while sources 3b, 4b and 5b are barely or even not detected in the F150W+F200W+F277W stacked image (probing the ultraviolet wavelengths 1600\AA-2800\AA) and are referred in the main text as nebular knots possibly hosting dust-obscured star clusters. 
 The physical quantities of the SED fitting are reported in Table~\ref{tab:summary}.}
 \label{fig:SEDs}
\end{figure*}
\begin{figure}
\center
 \includegraphics[width=\columnwidth]{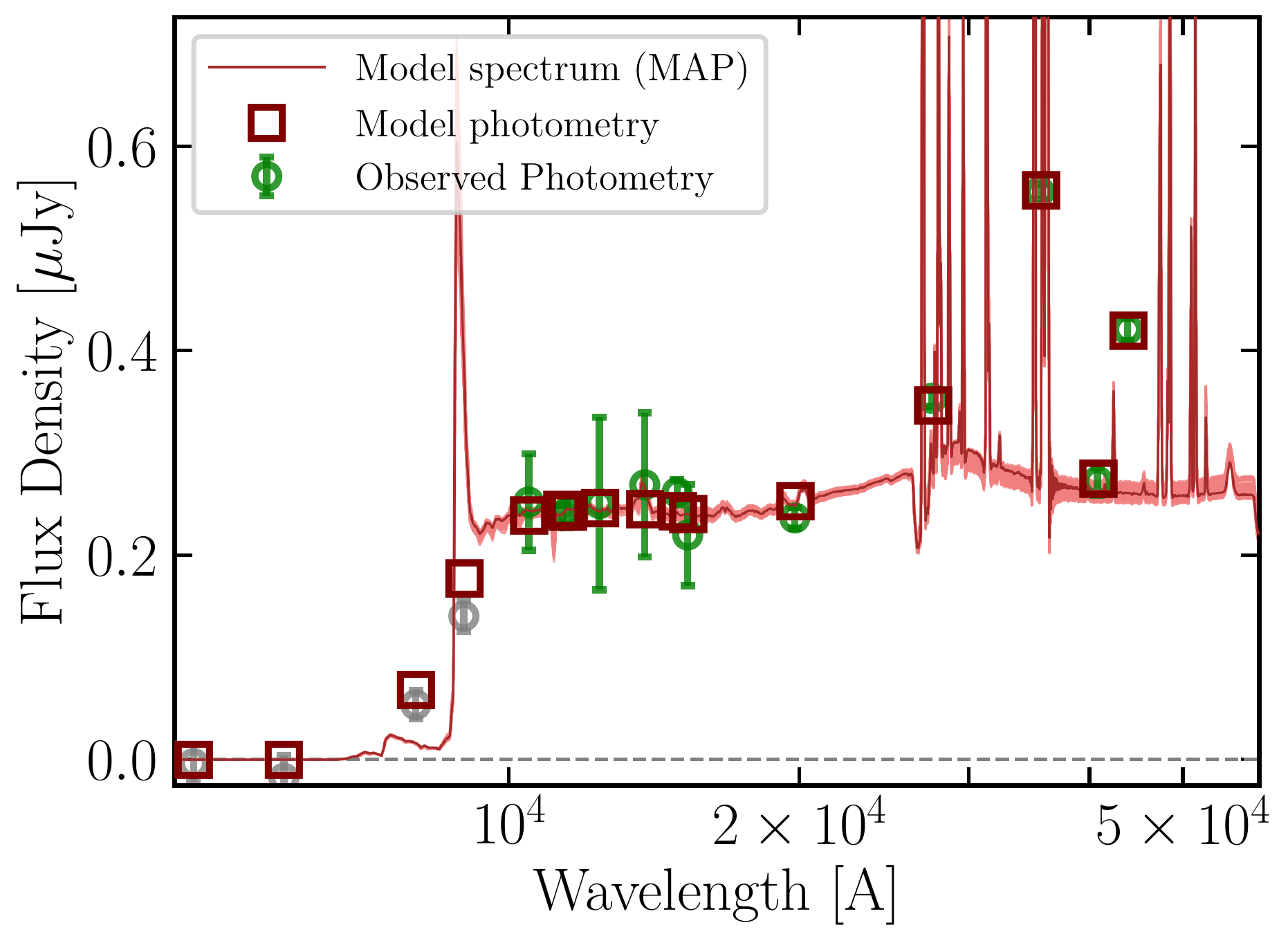}
 \caption{Prospector SED fit to lensed image d of the star-forming complex SFC (SFCd). The best-fit SED is shown in red, the fitted photometry in green (grey points are masked photometry values), and the best-fit model photometry is depicted in brown. 
 The physical quantities inferred from the SED fitting process are reported in Table~\ref{tab:summary}.}
 \label{fig:SFC}
\end{figure}

The SED fitting analysis has been performed following the methodology described in \citet{claeyssens22}. 

Motivated by the small physical scales probed ($\lesssim 10\,{\rm pc}$), which easily overlap with star-forming regions and star clusters sizes in the local universe, we fit with the single stellar population models Yggdrasil \citep[][]{Zackrisson2011}. These models include a full self-consistent treatment of ionized gas emission using Cloudy \citep{Ferland1998}, with a fully populated \cite{Kroupa2001} IMF. The model spectra have been redshifted to $z=6.0$, and \cite{Calzetti2000} dust attenuation has been applied. We adopted two different metallicities $Z = 0.004$ and $Z = 0.0004$ (corresponding to about 30\% and 3\% of the solar metallicity as determined by \citealt{Asplund2009})  
finding both fit the observed SEDs similarly well (Figure~\ref{fig:SEDs}). It is known that optical SED rest-frames are not very sensitive to metallicity, therefore we report in Table~\ref{tab:summary} the best solutions using models with 30\% solar abundances, but we include also the best-fitted model with 3\% solar metallicity in Figure~\ref{fig:SEDs}. We note the solutions determined with 3\% solar metallicity yield physical parameters in agreement within the uncertainties. The associated uncertainties have been derived using the 68\% distributions recovered by fitting 100 realizations of the observed SEDs after photometric uncertainties have been propagated using Monte Carlo sampling. 

Four of the six compact cluster candidates, 3b, 4b, 5b, and 6b, have clear photometric jumps in the broadbands containing strong emission lines, which imply large equivalent widths of the optical nebular lines. Conversely, source 1b and 2b do not show such a prominent nebular emission in the photometric SEDs, implying a smaller ionization field and more evolved stellar populations.  The modeling of the SEDs shown in Figure~\ref{fig:SEDs} and~\ref{fig:SFC} illustrates this behavior. The resulting ages reported in Table~\ref{tab:summary} confirm the expectations. The ages of the clusters become progressively younger along the arc, starting at 30 Myr down to 1 Myr. Recovered masses are all above $10^6$~\msun, in agreement with the completeness limits for point sources described above.

\subsection{Young massive star clusters populating the {\tt Sunrise} galaxy}
\label{starclusters}

The very small sizes ($<10$ pc) and the large stellar mass surface densities ($> 1000$~M$_{\odot}$~pc$^{-2}$) 
are fully consistent with those observed in young massive ($>10^5~$M$_{\odot}$) stellar clusters \citep[e.g.,][]{Brown2021} and globular clusters \citep[e.g.,][]{norris2014} in the local Universe. Furthermore, the inferred dynamical age $\Pi$, defined as $\Pi$ = age / T$_{CR}$  (with T$_{CR}$ the crossing time, \citealt{gieles11, adamo20}),\footnote{$\Pi$ has been extensively used to identify bound star clusters in the local Universe \citep[e.g.,][]{adamo20extreme,ryon17}.
The crossing time expressed in Myr is defined as ${\rm T_{CR} = 10 \times (R_{\tt eff}^{3} / GM)^{0.5}}$, where M and ${\rm R_{eff}}$ are the stellar mass and the effective radius, respectively, and ${\rm G \approx 0.0045~pc^{-3} \msun^{-1} Myr^{-2}}$ is the gravitational constant.} implies that the majority of the sources are gravitationally bound stellar clusters ($\Pi>1$, see Table~\ref{tab:summary}). Two of them, 1b and 2b appear relatively evolved (with no evidence of detectable ionized gas emission in their observed SEDs, Figure~\ref{fig:SEDs}), with ages spanning the range $\simeq$ 10-30 Myr and $\Pi>>1$. Remarkably, both of them show properties reminiscent of local massive globular clusters (see Sect.~\ref{sect:GC}). On the other hand, cluster candidate 6b is in the bursty formation phase with an age $<5$ Myr and still appears as a bound young star cluster (although note that $\Pi$ at such young ages is ill-defined, with the crossing time comparable to the age of the system), in which the strong ionization field provided by hot stars pump significantly large optical emission lines (see Sect.~\ref{sect:xi}).

It is worth noting the presence of three additional knots (3b, 4b and 5b) which are clearly visible in the F277W, F356W and F444W bands only, without any clear $-$ yet compact $-$ counterpart in the stacked image (F115W + F150W + F200W), which covers the rest-UV continuum (see Figure~\ref{fig:summary}). Interestingly, the bands where the sources are detected include the \oiidoub, \hb+\oiiidoub\ and \ha\ lines, respectively. For this reason we refer to them as {\it nebular dominated knots}. The sizes of these nebular knots have been inferred from the F277W band which provides the sharpest PSF (among the three mentioned above) and are remarkably small, with effective radii spanning the interval $R_{\rm eff}=4-25$ pc (although these values have large uncertainties).
Under the assumption that such rest-frame optical high-ionization lines, like the \oiiidoub, trace the location of hot stars (as observed in the local Universe, e.g., \citealt{james16, sirressi22_HARO11}), it is plausible that these detected compact nebular emissions host stellar clusters. Their SED analysis suggests that these are hidden clusters. They have significant dust attenuation (E(B$-$V) between 0.15 and 0.35 mag), which explains why we do not clearly detect them in the UV, while at longer (optical rest-frame) wavelengths emerge by means of prominent emission lines. This is the first evidence of star clusters still {\it embedded} in their natal formation region detected at cosmological a distance. 

\section{Physical properties of the extended star-forming complex and host}

\subsection{Photometry of extended regions}
\label{app_isophot}

We extract isophotal photometry in nine regions detected along the arc.
We construct a detection image, created from background-subtracted F277W, F356W and F410M NIRCam mosaics. The area of interest shown in Figure \ref{fig:summary} is crowded with galaxy cluster members, background objects and the diffuse intracluster light, which presents a challenge toward robustly segmenting our sources. To ensure consistency between our detection parameters, and cover the large dynamic range of brightness of sources in the field, we separate the detection image to cover three regions along the arc, which contain the following knots - 1) 1d, 2d and 6d; 2) 6c, 6b, 2b and 1b; 3) 1a and 6a. We smooth the detection images with a median filter of \texttt{pixel\_size=2}, similarly to the procedure adopted in \citet{livermore17} and \citet{kokorev22}. Finally, the segmentation maps are produced by running \textsc{sep} \citep{sep}, and are further smoothed with a median filter of \texttt{pixel\_size=8}. The map of segments is shown in Figure~\ref{fig:summary}. 

We apply the final segments to all available \JWST\, and $HST$ mosaics, which we have matched to the F444W PSF. 
We also subtract the local background in each sub-region containing the knots. We designed our isophotal apertures to be fairly large, so we do not perform any further aperture corrections. Coordinates and photometry for objects analyzed in this work are given in Table~\ref{tab:seg_photometry}. 

\begin{deluxetable*}{ccccccccc}
\tablenum{2}
\tablecaption{\JWST/NIRCam isophotal photometry of the extended regions. \label{tab:seg_photometry}}
\tablewidth{0pt}
\tablehead{
\colhead{ID} & 
\colhead{F090W} & 
\colhead{F115W} & 
\colhead{F150W} & 
\colhead{F200W} & 
\colhead{F277W} & 
\colhead{F356W} & 
\colhead{F410M} & 
\colhead{F444W} 
}
\decimalcolnumbers
\startdata
SFCd &  26.03(0.12) & 25.44(0.07) & 25.36(0.05) & 25.47(0.05) & 25.03(0.02) & 24.54(0.01) & 25.31(0.05) & 24.84(0.03)  \\
TOTd & 25.54(0.10) & 25.12(0.07) & 25.07(0.05) & 25.10(0.05) & 24.68(0.02) & 24.29(0.01) & 24.86(0.05) & 24.54(0.03) \\
TOT & 24.28(0.06) & 23.71(0.03) & 23.74(0.03) & 23.69(0.02) & 23.26(0.01) & 22.86(0.01) & 23.35(0.02) & 23.11(0.01) \\
\hline
\hline
\enddata
\tablecomments{Magnitudes and 1-sigma errors are reported in each column for the extended regions reported in Figure~\ref{fig:summary}, as derived in Sect.~\ref{app_isophot}. The ID = ``TOT'' is the photometry of the full arc (which combines the three regions marked with transparent ellipses in Figure~\ref{fig:ratios}), and the SFCb and TOTd are reported in Figure~\ref{fig:summary}.
}
\end{deluxetable*}

\subsection{SED fitting analysis}

While the star cluster candidates have been fitted with instantaneous burst models (i.e., single stellar populations), 
we allow for an extended period of 
constant and flexible star formation in fitting the SFC,
appropriate for large regions of a galaxy, as well as the total flux from the host galaxy.

\subsubsection{BAGPIPES}

We perform SED fitting with \textsc{BAGPIPES} \citep{Carnall2018}
in a manner similar to \cite{Hsiao2022}.
Briefly, we use BPASS v2.2.1 SED templates \citep{BPASS}
reprocessed with \textsc{Cloudy} c17.03 
\citep{Ferland1998, Ferland2013, Ferland2017}.
We use the fiducial BPASS IMF {\tt imf135\_300}
with a high-mass slope similar to \cite{Kroupa2002}.
We allow the ionization parameter $U$ to range between $\log(U) = -4$ to $-1$.
For dust attenuation, we use the \cite{Salim18} parameterization
capable of reproducing Milky Way and SMC attenuation curves.
Young stars (age $< 10$ Myr) residing in stellar birth clouds
experience more dust extinction by a factor $\eta$ in the range 1 to 3. 

For SFC, we assumed a constant star formation rate with one or two components.
The single-component result was
$\sim$2 \msunyr\ over $\sim$20 Myr 
after delensing (Table \ref{tab:summary}).
The two-component result was 
$\sim$ 3\msunyr\  over $\sim$10 Myr
plus a longer duration
$\sim$$0.1 M_\odot$ yr$^{-1}$ over $\sim$600 Myr.
The long-duration SFR result is similar to the result we obtain 
when fitting to the extended portion of the arc with no clumps.
Allowing for this longer duration component 
increases our mass estimate for SFC to $\sim$$10^8$\msun,
twice that derived when assuming a single component constant SFR.

We also explored SED fitting with 2016 version BC03 models \citep{BC03}
including empirical MILES stellar spectra \citep{miles_library,miles_2016},
similarly reprocessed with \textsc{Cloudy}.
These templates failed to reproduce the strong nebular emission lines inferred in SFC
(see \citealt{Wofford16} for a similar result).
In BPASS models, binary interactions produce helium stars 
that are hotter ($T_{\rm eff} = 10^5$ K, c.f.\ \citealt{Goetberg2018,Goetberg2019}) yielding stronger O32 line ratios \citep{Lin18,Lin19}.

\subsubsection{Prospector}

We perform complementary SED fitting analyses of observed photometry in the star-forming clump (SFC) and the host galaxy (TOTd) with the Markov Chain Monte Carlo (MCMC)—based stellar population synthesis (SPS) and parameter inference code, {\tt Prospector}. 

{\tt Prospector} is based on the Python-FSPS framework, with the MILES stellar spectral library and the MIST set of isochrones (\citealp{Conroy_2010,prospector,miles_library,mist_library}). 

In these models, we fit for a non-parametric star formation history observed at a notional redshift $z=6$. We utilize age bins with $[0-20]$, $[20-50]$, $[50-100]$, $[100-300]$, $[300-600]$ and $[600-900]$ Myr in lookback time), represented by the parameters SFR$_{\rm ratio}$, referring to the ratio of total star formation in adjacent time bins. 
The priors for the star-formation rate ratios in adjacent time-bins ascribe to the continuity prior. These parameters fit for the change in $\log({\rm SFR})$ between the bins, and statistically prefer smooth transitions in SFR (see \citealp{Leja_2019}). 

We also fit for dust attenuation using the \cite{Kriek_2013} attenuation law applied to all the light from the galaxy (in units of opacity at 5500\AA), while simultaneously constraining the index of the attenuation power law. We also use as free parameters the 
stellar and gas-phase metallicities $\log(Z/Z_{\odot})$ (where $Z_{\odot}=0.0142$ as determined by \citealt{Asplund2009}, and flat priors between [$-2.0$,$0.2$]), gas ionization parameter $U$ (with flat priors sampled between [$-4$,$-1$] similar to the BAGPIPES analysis), and the remnant stellar mass in the galaxy (M$_{\rm remnant}$, in units of M$_{\odot}$), as free parameters.
We assume a Chabrier IMF \citep{Chabrier_2003}, and IGM absorption is present in these models. 
Nebular continuum and line emission are used in the fit, and a nominal velocity smoothing of 200 km s$^{-1}$ was used for the model spectrum. 
For more details on applying non-parametric star formation history models to lensed galaxy observations that sample rest-frame UV and optical emission, refer to, e.g.,  \cite{Khullar2021} and \cite{Sukay2022}.

We mask photometry redwards of F090W in this analysis, which samples Ly$\alpha$ emission (that is not constrained without spectroscopy, and is decoupled from star formation due to geometrical considerations).
The observed photometry from one of SFC's images, best-fit SED model and predicted model photometry are shown in Figure \ref{fig:SFC}.
We also provide the inferred parameter values -- including remnant (delensed) stellar masses and mass-weighted ages -- in Table \ref{tab:summary}.

\subsection{Mapping the SFH of the {\tt Sunrise} arc}

The SFH model assumptions need to be incorporated when inferring the remnant stellar mass, ages and the mass assembly in the Sunrise arc (and its constituents). After considering uncertainties ($\sim0.4$\,dex in both analyses) we infer a 1.4\,$\sigma$ difference between the {\tt BAGPIPES} (constant SFH, $\log {\rm M}_{\rm median}\sim8.5$) and {\tt Prospector} stellar masses (flexible SFH with 6 age bins, $\log {\rm M}_{\rm median}\sim9.3$). 

We attribute this difference to the modeling assumptions, e.g., \cite{Leja_2019} find a difference of 0.2-0.4 dex in median values of stellar mass between inferred values from parametric vs non-parametric star formation histories. For photometric fitting, this also has consequences for inferred mass-weighted ages, metallicities and dust properties. For the total galaxy flux and SFC, our {\tt Prospector} analysis results in a mass-weighted age of $\sim$ 200 Myr and $\sim$ 140 Myr respectively; this is an older inferred age than the single stellar population (SSP) or constant star formation history modeling.

\begin{deluxetable*}{lcccccccccc}
\tablenum{3}
\tablecaption{Intrinsic (source plane) physical properties of the multiple image systems shown in Figure~\ref{fig:pano}. \label{tab:summary}}
\tablewidth{0pt}
\tablehead{
\colhead{ID} & 
\colhead{M$_{\rm 2000}$} & 
\colhead{M$_{\rm 5700}$} & 
\colhead{Stellar Mass} & 
\colhead{Age} & 
\colhead{E(B-V)} & 
\colhead{R$_{\rm eff}$} & 
\colhead{$\Pi$} & 
\colhead{$\Sigma_{\rm Mass}$} & 
\colhead{$\mu_{\rm tot}$} & 
\colhead{$\mu_{\rm tang}$} \\
\colhead{} 
& \colhead{F150W} 
& \colhead{F410M} 
& \colhead{[$10^{6}$~\msun]} 
& \colhead{[Myr]} 
& \colhead{}
& \colhead{[pc]~[mas]} 
& \colhead{} 
& \colhead{[$10^{3}$~\msun~pc$^{-2}$]} 
& \colhead{}  
& \colhead{}
}
\decimalcolnumbers
\startdata
1b  &  -15.22  &  -15.54  &  $7.1^{+1.6}_{-4.6}$  &  $30_{-22}^{+0}$  & $0.00_{-0.00}^{+0.10}$  &  $1.4_{-0.7}^{+0.3}~[15]$ &  $314.1_{-155.8}^{+297.6}$  & $311.3_{-158.1}^{+445.5}$ & $\gtrsim$66  &  $\gtrsim$60\\ 
2b  &  -15.34  &  -15.41  &  $3.9^{+0.2}_{-1.3}$  &  $10_{-1}^{+0}$  & $0.10_{-0.05}^{+0.00}$  &  $6.3_{-1.3}^{+1.1}~[30]$ &  $8.3_{-1.9}^{+3.1}$  & $8.7_{-2.7}^{+4.8}$ & $\gtrsim$30  &  $\gtrsim$27\\ 
3b$^{\star}$  &  -13.58  &  -14.26  &  $1.1^{+8.7}_{-0.5}$  &  $4_{-3}^{+36}$  & $0.25_{-0.20}^{+0.20}$  &  $6.1_{-3.6}^{+12.5}~[29]$ &  $1.9_{-0.9}^{+15.7}$  & $2.7_{-2.3}^{+16.3}$ & $\gtrsim$30  &  $\gtrsim$27\\ 
4b$^{\star}$  &  -16.84  &  -16.91  &  $10.1^{+11.0}_{-0.2}$  &  $1_{-0}^{+3}$  & $0.15_{--0.05}^{+0.15}$  &  $24.8_{-12.3}^{+62.6}~[117]$ &  $0.2_{-0.1}^{+0.4}$  & $1.5_{-0.8}^{+2.1}$ & $\gtrsim$30  &  $\gtrsim$27\\ 
5b$^{\star}$  &  -13.50  &  -14.51  &  $3.1^{+10.2}_{-2.0}$  &  $6_{-5}^{+74}$  & $0.40_{-0.30}^{+0.25}$  &  $4.9_{-1.7}^{+10.6}~[23]$ &  $6.6_{-3.3}^{+62.6}$  & $11.8_{-9.1}^{+41.6}$ & $\gtrsim$30  &  $\gtrsim$27\\ 
6b  &  -15.70  &  -15.83  &  $3.3^{+3.2}_{-0.8}$  &  $4_{-3}^{+2}$  & $0.15_{-0.10}^{+0.05}$  &  $8.5_{-3.0}^{+2.1}~[40]$ &  $2.0_{-1.1}^{+2.2}$  & $4.1_{-2.4}^{+5.6}$ & $\gtrsim$30  &  $\gtrsim$27\\ 
\hline
SFCd$^{+}$ & -17.64  & -17.69  &  $266 ^{+234}_{-122}$  &  $140 ^{+140}_{-80}$  & $0.24^{+11}_{-10}$  & $\simeq 200^{\dag}$  & $-$  & $-$ & 18  & 16 \\
SFCd$^{\rm o}$  & -17.64  & -17.69  &  $44\pm11$  &  $11\pm2$    & $-$  & $\simeq 200^{\dag}$  & $-$  & $-$ &18  & 16 \\
SFCd$^{\rm oo}$ & -17.64  & -17.69  &  $97\pm20$  &  $130\pm35$  & $-$  & $\simeq 200^{\dag}$  & $-$  & $-$ &18  & 16 \\
\hline
TOTd$^{+}$ & -17.93  & -18.14 & $2200^{+3010}_{-1200}$ & $201^{+93}_{-108}$ & $0.09^{+0.05}_{-0.04}$ & $\simeq 1000^{\dag}$ & $-$  & $-$ & 15 & 13\\
TOTd$^{\rm oo}$  &-17.93 & -18.14  &  $300^{+60}_{-40}$  &  $185\pm50$  & $-$  & $-$ $\simeq 1000^{\dag}$ & $-$ & $-$ & 15 & 13 \\
\hline
\hline
\enddata
\tablecomments{Column $\#1$ lists the IDs of the sources; columns $\#2$ and $\#3$ report the absolute magnitudes in the rest-frame ultraviolet and optical. Such magnitudes are likely upper limits given the limits on magnification (see column \#10); in column $\#4$, $\#5$ and $\#6$ show the stellar masses, ages and E(B-V) inferred from the SED fitting; the effective radius is listed in $\#7$, based on F150W band for 1b, 2b, 6b and SFC, and F277W for the nebular knots 3b, 4b and 5b; columns $\#8$ and $\#9$ summarize the dynamical age $\Pi$ and the stellar mass surface density $\Sigma_{Mass}$. Lower limits on the total and tangential magnifications are reported in column $\#10$ and $\#11$, respectively.
In the approximation in which $\mu_{tot} \simeq \mu_{tang}$ as in this system, the stellar masses and the effective radii scale as $\mu_{tot}^{-1}$ and $\mu_{tang}^{-1}$, whereas $\Pi$ and $\Sigma_{Mass}$ are directly proportional to $\mu_{tot}$.}
\tablenotetext{\star}{nebular knots.}
\tablenotetext{+}{~Prospector-based SED fitting with flexible star-formation-histories. Quoted ages are mass-weighted.}
\tablenotetext{\rm o}{~BAGPIPES-based SED fitting with one (o) and two (oo) components of constant star formation. Quoted ages are mass-weighted.}
\tablenotetext{$\dag$}{The size refers to the extent of the region (not the effective radius). The reported error bars do not include uncertainties on the magnifications which are considered lower limits in this work (see Appendix~\ref{app_photoz}).}
\end{deluxetable*}

\section{Discussion}
\label{sec:discussion}

\subsection{Formation of proto-globular clusters in the {\tt Sunrise} arc?}
\label{sect:GC}

Since we have access to the host's physical properties, we can provide first-order estimates of the total cluster formation efficiency (CFE) in the {\tt Sunrise} galaxy (TOTd). CFE is broadly defined in the literature as the ratio between the total mass forming in stars divided by the total stellar mass formed in the region within a given interval of time \citep[see][]{krumholz2019}. Following \citet{adamo17}, the CFE is the ratio between the {\it cluster formation rate} and the star formation rate of the galaxy hosting the star clusters, both calculated over a comparable interval of time $dt$ (with the cluster formation rate defined as the total stellar mass forming in cluster over the same $dt$).
The {\tt BAGPIPES} and {\tt Prospector} analyses suggest the galaxy has built 
between $\sim 3 \times 10^8\,M_\odot$ and $\sim 2 \times 10^9\,M_\odot$ stellar mass
over the course of 200\,Myr or so. 
If we assume these to be lower and upper mass estimates,
we can compare them with the total mass in star clusters by dividing each mass estimate by their respective ages, to get a first-order estimate of the SFR. The 
total mass in the cluster is $3.0\times 10^{7}$ \msun\ formed over 30 Myr, resulting in an SFR of 1~\msunyr\ for the clusters. In the case of the host, the SED fit analyses find a minimum and maximum 
SFR of 3 up to 10~\msunyr. The ratio will result in a CFE between 10 and 30~\% for the entire {\tt 
Sunrise} arc. These estimates do not account for the mass in star clusters below the detection limits, so they should be considered lower-limits. It is interesting to notice that the recovered values bracket the CFE reported for the {\tt Sunburst} arc at redshift 2.4 by \citet{vanz22_CFE}.

The area of the segment that contains the whole host (TOTd) corresponds in physical scales to $6.0\pm2.0$ kpc$^2$ (adopting an error of 30~\% on the magnification). The total minimum and maximum SFR surface density ($\Sigma_{\rm SFR}$) are therefore $0.5$ to $1.7$ \msunyr~kpc$^{-2}$. Extrapolating from the cluster formation efficiency vs. $\Sigma_{\rm SFR}$ relation \citep[e.g.][]{adamo20extreme} we expect about 30 to 60\% of the star formation in the host to happen in YMCs, which is in agreement with the lower-limits reported above. 

 These high cluster formation efficiencies are similar to those reported in the local universe for green-pea analogs at distances $\sim$50 Mpc, namely, ESO338-IG04 and SBS0335-052E \citep{adamo2011}. Both these galaxies host very massive clusters ($>10^5$ \msun). ESO330-IG04 harbors a $\sim$ 6 Myr old compact cluster with measured dynamical mass of $1.3\times10^7$ \msun\ \citep{ostlin2007}. SBS0335-052E is among the most metal-poor ($\sim5$\% Z$_{\odot}$) galaxies in the nearby universe \citep{herenz17_ionized_channel}. It hosts 6 YMCs with masses well above $10^5$ \msun\ and ages younger than 15 Myr \citep{adamo2010}. The physical properties of the clusters hosted by SBS0335-052E show very similar age spread as observed in the YMCs of the {\tt Sunrise} arc. The EW(H$\alpha$) of the two youngest clusters in SBS0335-052E is a few-thousand~\AA, suggesting very young ages, similar to the knots embedded in the SFC complex.

Applying the conversion by \citet{vazdekis2015}\footnote{[Fe/H]$ = $[M/H]$ - A \times$ [$\alpha$/Fe], where $A=0.75$} and assuming an average [$\alpha$/Fe] value of 0.3 \citep{recioblanco2018}, we derive for 1b and 2b a [Fe/H] value lower-limit of $-1.7$ ($0.03$~Z$_{\odot}$) and upper-limit of $-0.7$ ($0.3$~Z$_{\odot}$). Their ages at redshift 6 correspond to a formation age of $\sim12.8$ Gyr. 

Comparing to the observed age-metallicity relation in the Local Group, where age determination for GCs is more robust, we observe that the {\tt Sunrise} arc YMCs have metallicity ranges that overlap with the old ($\sim$12 Gyr) metal-poor branch of GCs of the Milky Way and Large Magellanic Cloud \citep{forbes2010, narloch2022}.

The derived masses of the YMCs are all above $10^6$~\msun\ (Table~\ref{tab:summary}). Even by making the conservative assumption that about 75\% of their mass is lost during a Hubble time evolution \citep{reinacampos2018}, their masses will still be close to or larger than the characteristic mass of GC populations observed in the local universe (e.g., $\sim2 \times 10^5$ \msun, \citealt{brodie2006}). We conclude that the YMCs in the {\tt Sunrise} arc might evolve into metal-poor GCs that will likely populate the outskirts of galaxies at redshift zero.

 Finally, by simply dividing the total mass in clusters versus the host one, we find between 1 and 10 \% of the total stellar mass in the {\tt Sunrise} galaxy is located in these young star clusters. To understand whether the fractions we report are extreme we can look at the local universe for a comparison. However, young star cluster formation in the local universe is happening in already evolved disk galaxies, for which the bulk of the mass has been assembled around the cosmic noon. It is therefore more appropriate to look at globular cluster populations. \cite{larsen2012} report that the total stellar mass in the metal-poor globular clusters associated with the Milky Way halo is about 2\% of the total mass of the halo component. On the opposite extreme end, in the low-mass dwarf spheroidal Fornax, 5 massive and metal-poor GCs constitute about 30 to 50 \% of the total stellar mass in the galaxy at those metallicity ranges. So we conclude that the cluster population detected in the {\tt Sunrise} arc is well within the expectation of star cluster formation during galaxy assembly.      

\subsection{High ionizing photon production efficiency }
\label{sect:xi}
\begin{figure*}
\center
 \includegraphics[width=\textwidth]{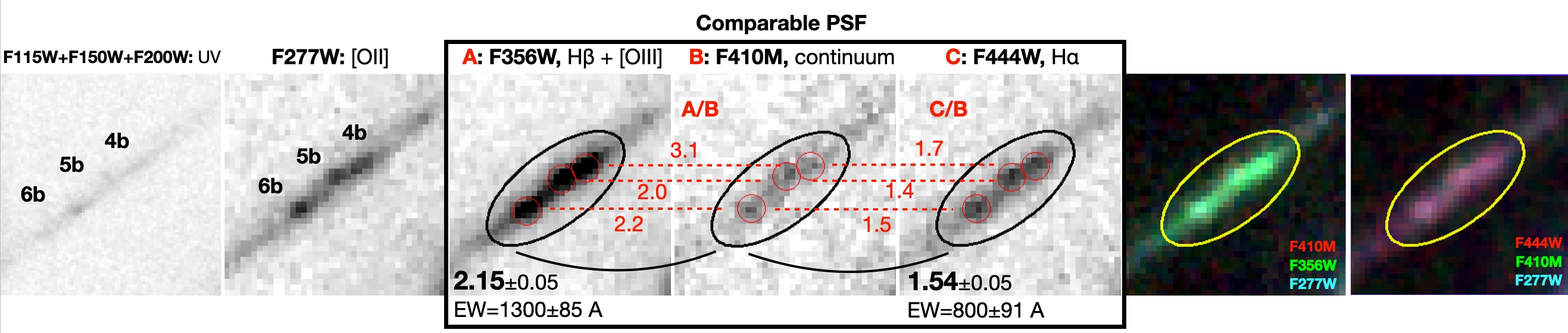}
 \caption{The star-forming complex (SFC) at position ``b''. From left to right: the stacked image probing the ultraviolet wavelength (1600-2800~\AA), the F277W in which knots 4b, 5b and 6b are identified, the F356W, F410M and F444W probing the indicated lines and continuum with reported the flux ratios (image A/ image B) for each knot (red) and for the elliptical aperture (black) corresponding to rest-frame equivalent widths EW~$\simeq1300$\AA\ and 800\AA\  for \hb+\oiiidoub\ and \ha\ nebular lines, respectively. The rightmost images show the same color images shown in Figure~\ref{fig:pano}.  }
 \label{fig:EWs}
\end{figure*}

A prominent feature of the {\tt Sunrise} galaxy that clearly emerges by means of our \JWST/NIRcam imaging is the multiple image star-forming complex SFC shown in Figures~\ref{fig:pano} and~\ref{fig:summary}. In such a region $-$ which extends about 200 parsecs in the source plane $-$ significant enhanced flux is detected in the F356W band if compared to the continuum probed by the F410M filter (corresponding to $\simeq 5.9\times10^{-17}$ \ergscm) potentially implying rest-frame equivalent widths of the group of lines \hb+\oiiidoub\ of $\simeq 1300$~\AA\ (Figure~\ref{fig:pano} and see Figure~\ref{fig:SEDs}). The same region also shows a clear excess in the F444W (producing a flux $\simeq 2.3\times10^{-17}$ \ergscm), ascribed to the strong \ha\ line with a rest-frame equivalent width $\simeq 800$~\AA. Figure~\ref{fig:EWs} shows the zoomed SFC at position ``b'' (the most magnified one) in which knots 4b, 5b and 6b are clearly identified, along with their significant flux contrast imprinted in the \JWST/NIRcam photometry (Figure~\ref{fig:SFC}), overall associated to large rest-frame equivalent widths.
As discussed above, SFC hosts very young stellar clusters ($\lesssim 10$ Myr; 4b, 5b and 6b, associated to a large specific star formations rate, sSFR $\simeq 100$ Gyr$^{-1}$), in which O-type stars produce a vigorous ionization field and strong lines boosting the photometry as observed (the fraction of ionizing photons converted into lines is unknown, we assume $>50\%$).
The SED fitting of SFC is shown in Figure~\ref{fig:SFC} and the properties reported in Table~\ref{tab:summary}.
The large specific star formation rate and the 
large EWs of the optical nebular lines also imply a large ionizing photon production efficiency $\xi_{ion}$, accordingly to the positive correlation with the \oiiidoub\ line strength \citep[e.g.,][]{chevallard18, emami2020_xi, nakajima2020}. Precise line fluxes will be measured with \JWST/NIRSpec, however, $\xi_{ion}$ can also be estimated with the present data. 
The \ha\ luminosity associated with the flux derived above and combined with the ultraviolet luminosity at 1500~\AA\ inferred from the F115W-band provides $\log(\xi_{ion}[\rm Hz~erg^{-1}]) \sim 25.7$ without any dust correction. 
Corrections for dust attenuation or leakage of ionizing photons could decrease or increase $\xi_{ion}$, respectively. 
Such a crude value is in line with previous studies at similar faint luminosity regimes and \ha\ equivalent widths \citep[e.g.,][]{nakajima2020, nakajima2022, maseda20}, and  with the values reported recently by \citet{matthee22_nircam_slitless} derived with \JWST/NIRCam slitless spectroscopy at $z\sim 5-7$.
While the direct measure of escaping ionizing radiation is not possible at the redshift of {\tt Sunrise}, it is however worth noting that the presence of multiple star clusters (young and relatively evolved) implies that the {\tt Sunrise} galaxy was subjected to a progressive injection of energy and stellar feedback in the interstellar medium (e.g., 1b and 2b were active a few tens Myr before SFC) eventually favoring the construction of ionized channels and the escape of ionizing radiation into the intergalactic medium. A direct similar example has been observed in the {\tt Sunburst} Lyman continuum galaxy at $z=2.37$ (\citealt{vanz22_CFE}; see also \citealt{rivera19}; \citealt{mainali22}). Also, in local starburst galaxies such conditions have been recently observed with great detail \citep[e.g.,][]{bik18, sirressi22_HARO11}. 
Such an iterated star cluster formation would make the escape of ionizing radiation an intermittent process, which temporally correlates with the bursty star-forming phases \citep[e.g.,][]{wise2014, trebitsch2017}.

\section{Concluding remarks}

The optical rest-frame coverage at $z \simeq 6.0$ (up to 7000~\AA) and the enhanced angular resolution (three times better than HST in the H-band) provided by \JWST/NIRCam have allowed us to derive the stellar masses, ages, dust attenuation and refine the sizes of the compact sources hosted in the {\tt Sunrise} arc, which we conclude are massive stellar clusters (see Table~\ref{tab:summary}). It is worth noting that if \JWST/NIRSpec confirms the presence of emission lines in these systems at a much higher spectral resolution, \JWST/NIRCAM imaging still provides a continuous two-dimensional view with a remarkably high spatial contrast (down to a few parsec scales along the tangential stretch) of such prominent lines along the arc. There are three main results from our study:

\noindent $\bullet$ {\it Proto-globular clusters in the reionization era.} The inferred properties of the clusters 1b and 2b (i.e., the effective radii $<5 \rm \ pc$, old dynamical ages $\Pi \gg 1$, and stellar mass surface densities well above $1000$~M$_{\odot}$~pc$^{-2}$) qualify them as bound star clusters.  
If they survive till the present day, these systems would appear as a few $\sim 10^{6}$ \msun, $\simeq 12.5-13$ Gyr old likely metal-poor globular clusters.  
Source 6b is also massive but still very young. It is part of a larger star-forming complex (SFC; $\sim 200$ pc size), which includes two additional nebular knots plausibly hosting relatively dense and UV dust-attenuated star clusters. Such systems might host globular cluster precursors caught in the act of their initial formation phase. 

\noindent $\bullet$ {\it Ionization.} The star-forming complex ``SFC'' hosts very young stellar populations (age $<5$ Myr)  and is likely located in star clusters sharing the same natal region (of $\sim 200$ pc size). SFC also dominates the nebular emission line budget of the {\tt Sunrise} galaxy. In particular, the ionizing photon production efficiency we estimate for SFC is remarkably high, $\log(\xi_{ion}[\rm Hz~erg^{-1}]) \sim 25.7$, which is fully in line with the large rest-frame equivalent widths of the optical nebular lines inferred from the photometric discontinuities (\oiiidoub~rest-frame equivalent widths $\gtrsim 1000$\AA). The high occurrence of such prominent optical nebular emission lines observed at $z>6$ \citep[e.g.,][]{boyett_2022a, matthee22_nircam_slitless} and the connection with bursty events driven by very young stellar clusters as probed in lensed galaxies such as {\tt Sunrise} (and similarly by the {\tt Sunburst} arc, \citealt{vanz22_CFE}) suggest the stellar cluster formation activity was significant during reionization.

\noindent $\bullet$ 
{\it Host-galaxy.} Mass-weighted ages estimates suggest that the stellar mass of the {\tt Sunrise} arc has been assembled in the last few hundred Myrs. Overall these six YMCs constitute less than 10 \% of the total mass of the host galaxy. When focusing on the recent star formation history, we estimate that the stellar mass in clusters is an important fraction of the recently formed stellar mass in the galaxy (lower-limits to the CFE are in the range 10-30~\%). We conclude that cluster formation currently taking place in the {\tt Sunrise} arc is an important contribution to the stellar mass build-up and galactic stellar feedback.

\section*{Acknowledgements}
This work is based on observations made with the NASA/ESA/CSA 
\textit{James Webb Space Telescope} (\JWST)
and \textit{Hubble Space Telescope} (\HST). 
These observations are associated with \JWST\ GO program 2282
and \HST\ GO programs 14096, 15842, and 16668.
The data were obtained from the Mikulski Archive for Space Telescopes (MAST) 
at the Space Telescope Science Institute (STScI), 
which is operated by the Association of Universities for Research in Astronomy (AURA), Inc., 
under NASA contract NAS 5-03127 for \JWST. 
We acknowledge financial support from NASA through grant JWST-GO-02282.

AA acknowledges support from the Swedish Research Council (Vetenskapsr\aa{}det project grants 2021-05559). 
AA and AC thank M. Messa for sharing an earlier version of his software.
EV acknowledges financial support through grants PRIN-MIUR 2017WSCC32, 2020SKSTHZ and INAF “main-stream” grants 1.05.01.86.20 and 1.05.01.86.31. 
We acknowledges support from
the INAF GO Grant 2022 “The revolution is around the corner: JWST will probe globular cluster precursors and Population III stellar clusters at cosmic dawn” (PI Vanzella).
J.M.D. acknowledges the support of projects and PGC2018-101814-B-100 and MDM-2017-0765.
MB acknowledges support from the Slovenian national research agency ARRS through grant N1-0238. 
PD acknowledges support from the NWO grant 016.VIDI.189.162 (``ODIN") and the European Commission's and University of Groningen's CO-FUND Rosalind Franklin program.
MO acknowledges support from JSPS KAKENHI Grant Numbers JP22H01260, JP20H05856, JP20H00181.
SF acknowledges support from NASA through the NASA Hubble Fellowship grant HST-HF2-51505.001-A awarded by the Space Telescope Science Institute, which is operated by the Association of Universities for Research in Astronomy, Incorporated, under NASA contract NAS5-26555. 
BW acknowledges support from NASA under award number 80GSFC21M0002.
LJF and AZ acknowledge support by Grant No. 2020750 from the United States-Israel Binational Science Foundation (BSF) and Grant No. 2109066 from the United States National Science Foundation (NSF), and by the Ministry of Science \& Technology, Israel.

%

\vspace{5mm}
\facilities{This work is based on observations made with the NASA/ESA/CSA 
\textit{James Webb Space Telescope} (\JWST)
and \textit{Hubble Space Telescope} (\HST).}





\appendix

\section{Photometric redshift of the {\tt Sunrise} arc}
\label{app_photoz}

\begin{figure}
\center
 \includegraphics[width=\columnwidth]{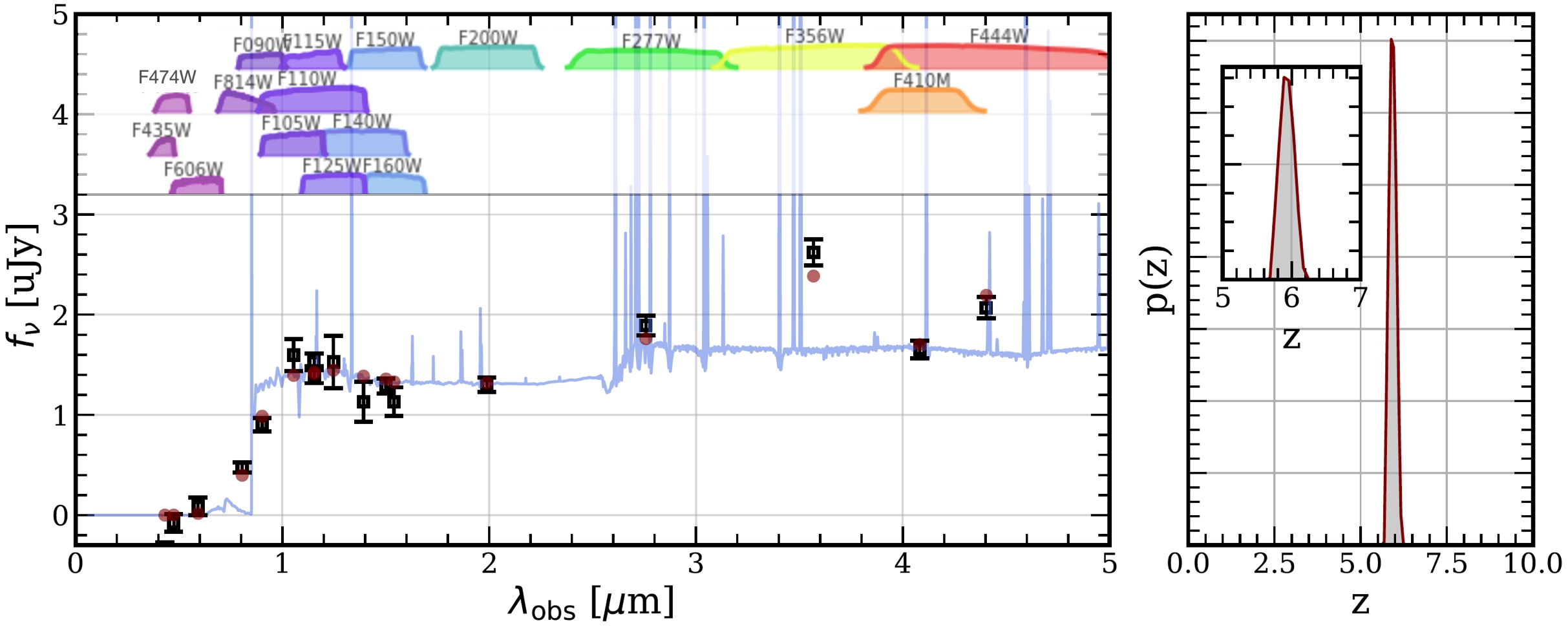}
 \caption{{\it Left}:
 Photometry of the full Sunrise Arc (black squares with error bars)
 and EAZY SED fit (blue spectrum; red circles for model fluxes).
 We sum PSF-matched isophotal photometry in all detected segments
 except 6c, which is blended with the interloper 
 (likely galaxy cluster member, see Figure~\ref{fig:summary}).
 {\it Right}: 
 Redshift probability distribution $z = 5.9^{+0.3}_{-0.2}$ (95\% CL). 
 }
 \label{fig:zphot}
\end{figure}

The extended wavelength coverage provided by \JWST/NIRCam imaging (up to 7000\AA~rest-frame) allows us to properly highlight the presence of underlying prominent optical nebular emission lines.
The color images reported in Figure~\ref{fig:pano} clearly show the enhanced flux produced by \hb+\oiiidoub\ in the F356W-band,
while the F277W-band (blue channel) and F410M-band (red channel) probe the underlying stellar continuum at 3800~\AA\ and 5700~\AA\ rest-frame without intercepting any significant emission lines, respectively.\footnote{It is worth noting that the F277W band includes the \oiidoub\ emission, whose effect, however, appears to be not significant in the present case.} 
Similarly, the same figure shows the \ha\ boosting in F444W (red channel). It is also worth noting that this new \JWST/NIRCam rest-frame optical view of the {\tt Sunrise} arc significantly improves the photometric redshift of the arc. In fact, the absence of the above lines in the medium filter F410M implies a minimum and maximum redshift of $\sim 5.5$ and $\sim 6.7$, which coupled with the $>2$ magnitudes drop at $\lambda < 0.9 \mu m$ places the arc at zphot~$=5.9^{+0.3}_{-0.2}$ without lower redshift solutions. The photometry of the full arc is used in the fit is obtained by combining all the segments as described in Figure~\ref{fig:summary}. The redshift probability distribution is shown in Figure~\ref{fig:zphot}.

\section{Magnifications of the star-forming regions}
\label{app_magnif}

\begin{figure*}
\center
 \includegraphics[width=\textwidth]{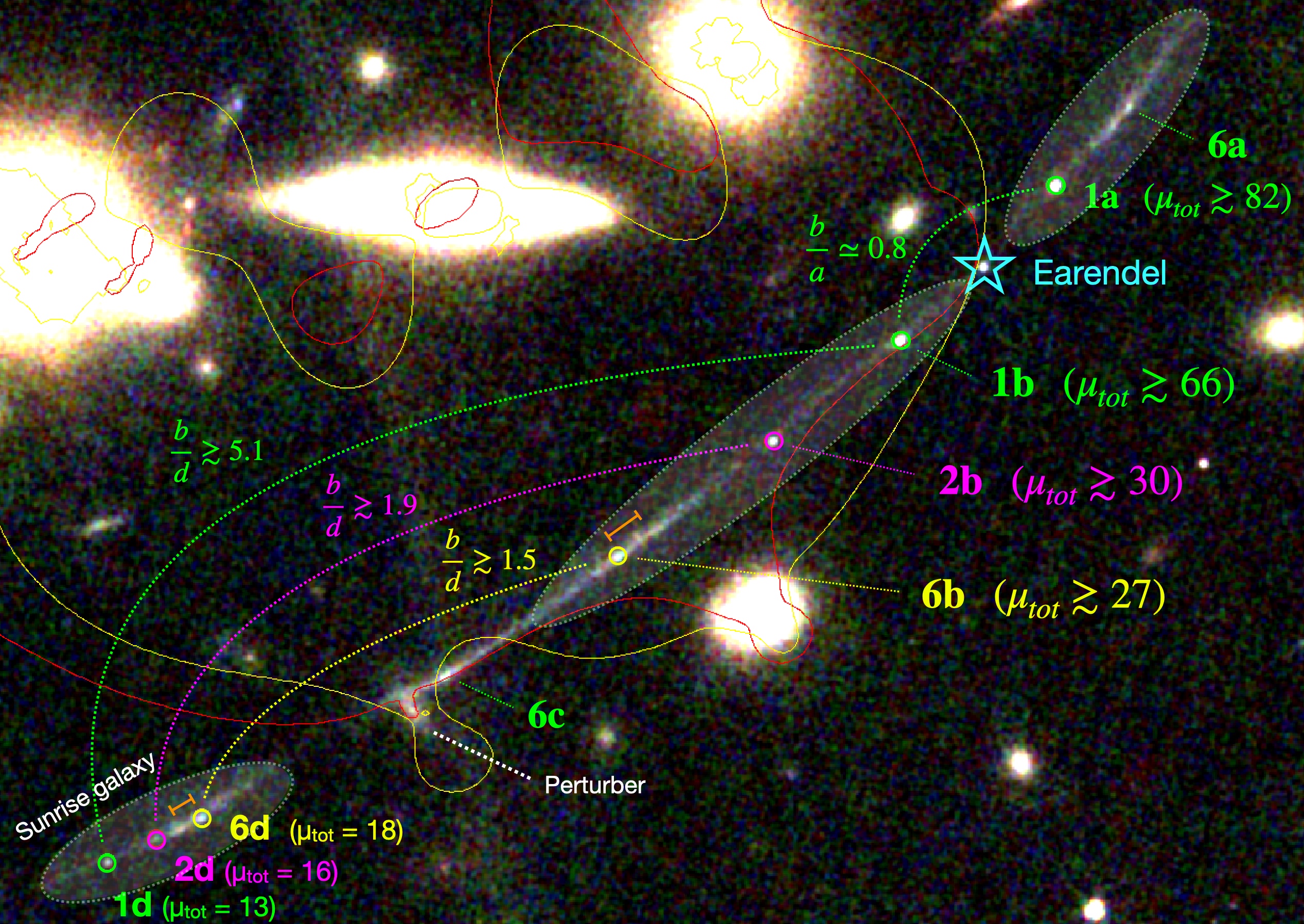}
 \caption{\JWST/NIRCAM color composites of the {\tt Sunrise} arc with indicated the knots detected in the ultraviolet wavelengths, 1 (green), 2 (magenta) and 6 (yellow), see Table~\ref{tab:summary}. Multiple image pairs are indicated with dotted lines adopting the same color coding. The flux ratios (``b" / ``d") are also reported for each pair. The amplifications $\mu_{tot}$ in "b" are indicated as likely lower limits (see text for details). The yellow and red contours show the critical curves for the current lens models based on JWST imaging (and will be presented in more detail in a forthcoming paper). While the configurations presented by the red curve trace more tightly the arc, the model presenting the yellow curve represents our most realistic and conservative amplification. The separations between knots in the SFC are indicated with orange segments at positions d and b, and their ratio of $\simeq 1.6$ (b/d) is in line with the expectation from the flux ratio reported. The three transparent white ellipses mark the same three multiple regions subjected to different magnifications. The region indicated with 6c (which is the same star-forming region observed on the other multiple segments, 6a,b,d) is associated with the perturber which bends the critical lines and amplifies a sub-component of the {\tt Sunrise} galaxy (a better analysis of that region will require spectroscopic confirmation and additional physical properties of the interloper and will be part of the work in preparation). }
 \label{fig:ratios}
\end{figure*}

Figure~\ref{fig:ratios} shows the \JWST/NIRCam color image of the compact sources detected in the ultraviolet ($1600-2800$~\AA) and reported in Table~\ref{tab:summary}. 

Lens models based on \HST\ imaging were presented and made publicly available by \cite{Welch22_nature}.
Significant updates will be possible
based on upcoming VLT/MUSE spectroscopic observations (PI.~G.~B.~Caminha). 
In this work, we make minor updates to lens models derived using three methods:
WSLAP+ \citep{Diego05wslap,Diego07wslap2};
Lenstool \citep{Jullo_Kneib_lenstool,Jullo_lenstool};
and GLAFIC \citep{Oguri_2010,Oguri_2021}.
The primary magnification estimates presented here are from WSLAP+. 
This model produces a critical curve that bends away from the arc after image 6c, resulting in magnifications for images 6d, 2d, and 1d lower than seen in other models. 
We adopt this as our baseline model to give the most conservative magnification estimate while the lens models are continuing to be refined. 
For comparison, we also show the preliminary updated Lenstool 
model, which produces critical curves that more tightly trace the arc (see red curve in Figure \ref{fig:ratios}), leading to higher predicted magnifications. This configuration introduces additional uncertainty on the magnification estimates, thus we adopt the lower magnification model to be conservative.

In this work we adopt a similar approach described in, e.g., \citet{vanz_paving}, to estimate the values of amplification for the most magnified knots (labeled as “b” in 
Figure~\ref{fig:ratios}). Specifically, the magnifications of the least magnified images (group “d”, which are affected by modest uncertainties, $<10\%$) are rescaled to ``b" locations accordingly to the measured flux ratios between “d” and “b” pairs. The flux ratios were inferred from the stacked image F115W+F150W+F200W. 
Unlike in the case of gravitationally lensed quasars, for strongly lensed compact stellar regions of sizes greater than 0.5 pc, magnification ratios are unaffected by microlensing effects and hence reliable proxies for the magnification ratio. Figure~\ref{fig:ratios} shows the locations of knots 1b,d, 2b,d and 6b,d and the inferred flux ratios measured within circular apertures of fixed diameter of  0\farcs2. The three knots (1, 2 and 6) bracket the entire portion of the {\tt Sunrise} arc we are analyzing in this work. It is worth noting that the physical regions probed by using fixed aperture sizes are different among multiple images because of the underlying different amplification. In particular, the larger physical regions probed by the apertures in the group ``d" (w.r.t. the group "b") imply the flux ratios are lower limits. Consequently, the rescaled values of $\mu_{tot}$ at ``b" reported in Figure~\ref{fig:ratios} and Table~\ref{tab:summary} are likely lower limits (after adopting $\mu$ values in ``d") and show values higher than 30. As an additional test, the orange segments show the separations between 4b$-$6b and 4d$-$6d whose ratio b/d~$\simeq 1.6$ (associated with the tangential magnification ratio) is consistent with the flux ratio reported (related to the total magnification ratio). 

Conservatively, we adopt all knots listed in Table~\ref{tab:summary} have $\mu_{tot}=30$, except 1b for which we adopt $\mu_{tot} \gtrsim 66$. The corresponding tangential magnification $\mu_{tang}$ along the arc is derived by dividing the above $\mu_{tot}$ values by the radial magnification, which in this case is quite stable along the full arc with a median value of $\simeq 1.1$.
The magnifications of the group ``d" are based on a new lens model which is based on \JWST/NIRCam identifications of the multiple images and will be presented in a forthcoming work. Table~\ref{tab:summary} reports such lower limits. 

\bibliography{sample631}{}

\begin{thebibliography}{}
\expandafter\ifx\csname natexlab\endcsname\relax\def\natexlab#1{#1}\fi
\providecommand{\url}[1]{\href{#1}{#1}}
\providecommand{\dodoi}[1]{doi:~\href{http://doi.org/#1}{\nolinkurl{#1}}}
\providecommand{\doeprint}[1]{\href{http://ascl.net/#1}{\nolinkurl{http://ascl.net/#1}}}
\providecommand{\doarXiv}[1]{\href{https://arxiv.org/abs/#1}{\nolinkurl{https://arxiv.org/abs/#1}}}

\bibitem[{{Adamo} {et~al.}(2011){Adamo}, {{\"O}stlin}, \&
  {Zackrisson}}]{adamo2011}
{Adamo}, A., {{\"O}stlin}, G., \& {Zackrisson}, E. 2011, \mnras, 417, 1904,
  \dodoi{10.1111/j.1365-2966.2011.19377.x}

\bibitem[{{Adamo} {et~al.}(2010){Adamo}, {Zackrisson}, {{\"O}stlin}, \&
  {Hayes}}]{adamo2010}
{Adamo}, A., {Zackrisson}, E., {{\"O}stlin}, G., \& {Hayes}, M. 2010, \apj,
  725, 1620, \dodoi{10.1088/0004-637X/725/2/1620}

\bibitem[{{Adamo} {et~al.}(2017){Adamo}, {Ryon}, {Messa}, {Kim}, {Grasha},
  {Cook}, {Calzetti}, {Lee}, {Whitmore}, {Elmegreen}, {Ubeda}, {Smith},
  {Bright}, {Runnholm}, {Andrews}, {Fumagalli}, {Gouliermis}, {Kahre}, {Nair},
  {Thilker}, {Walterbos}, {Wofford}, {Aloisi}, {Ashworth}, {Brown}, {Chandar},
  {Christian}, {Cignoni}, {Clayton}, {Dale}, {de Mink}, {Dobbs}, {Elmegreen},
  {Evans}, {Gallagher}, {Grebel}, {Herrero}, {Hunter}, {Johnson}, {Kennicutt},
  {Krumholz}, {Lennon}, {Levay}, {Martin}, {Nota}, {{\"O}stlin}, {Pellerin},
  {Prieto}, {Regan}, {Sabbi}, {Sacchi}, {Schaerer}, {Schiminovich}, {Shabani},
  {Tosi}, {Van Dyk}, \& {Zackrisson}}]{adamo17}
{Adamo}, A., {Ryon}, J.~E., {Messa}, M., {et~al.} 2017, \apj, 841, 131,
  \dodoi{10.3847/1538-4357/aa7132}

\bibitem[{{Adamo} {et~al.}(2020{\natexlab{a}}){Adamo}, {Zeidler}, {Kruijssen},
  {Chevance}, {Gieles}, {Calzetti}, {Charbonnel}, {Zinnecker}, \&
  {Krause}}]{adamo20}
{Adamo}, A., {Zeidler}, P., {Kruijssen}, J.~M.~D., {et~al.} 2020{\natexlab{a}},
  \ssr, 216, 69, \dodoi{10.1007/s11214-020-00690-x}

\bibitem[{{Adamo} {et~al.}(2020{\natexlab{b}}){Adamo}, {Hollyhead}, {Messa},
  {Ryon}, {Bajaj}, {Runnholm}, {Aalto}, {Calzetti}, {Gallagher}, {Hayes},
  {Kruijssen}, {K{\"o}nig}, {Larsen}, {Melinder}, {Sabbi}, {Smith}, \&
  {{\"O}stlin}}]{adamo20extreme}
{Adamo}, A., {Hollyhead}, K., {Messa}, M., {et~al.} 2020{\natexlab{b}}, arXiv
  e-prints, arXiv:2008.12794.
\newblock \doarXiv{2008.12794}

\bibitem[{{Asplund} {et~al.}(2009){Asplund}, {Grevesse}, {Sauval}, \&
  {Scott}}]{Asplund2009}
{Asplund}, M., {Grevesse}, N., {Sauval}, A.~J., \& {Scott}, P. 2009, \araa, 47,
  481, \dodoi{10.1146/annurev.astro.46.060407.145222}

\bibitem[{Barbary(2016)}]{sep}
Barbary, K. 2016, Journal of Open Source Software, 1, 58,
  \dodoi{10.21105/joss.00058}

\bibitem[{{Bergamini} {et~al.}(2022){Bergamini}, {Grillo}, {Rosati},
  {Vanzella}, {Mestric}, {Mercurio}, {Acebron}, {Caminha}, {Granata},
  {Meneghetti}, {Angora}, \& {Nonino}}]{bergamini_2022_J0416}
{Bergamini}, P., {Grillo}, C., {Rosati}, P., {et~al.} 2022, arXiv e-prints,
  arXiv:2208.14020.
\newblock \doarXiv{2208.14020}

\bibitem[{{Bik} {et~al.}(2018){Bik}, {{\"O}stlin}, {Menacho}, {Adamo}, {Hayes},
  {Herenz}, \& {Melinder}}]{bik18}
{Bik}, A., {{\"O}stlin}, G., {Menacho}, V., {et~al.} 2018, \aap, 619, A131,
  \dodoi{10.1051/0004-6361/201833916}

\bibitem[{{Bouwens} {et~al.}(2022){Bouwens}, {Illingworth}, {van Dokkum},
  {Oesch}, {Stefanon}, \& {Ribeiro}}]{bouwens2022_sizes}
{Bouwens}, R.~J., {Illingworth}, G.~D., {van Dokkum}, P.~G., {et~al.} 2022,
  \apj, 927, 81, \dodoi{10.3847/1538-4357/ac4791}

\bibitem[{{Bouwens} {et~al.}(2021){Bouwens}, {Illingworth}, {van Dokkum},
  {Ribeiro}, {Oesch}, \& {Stefanon}}]{bouwens21}
---. 2021, \aj, 162, 255, \dodoi{10.3847/1538-3881/abfda6}

\bibitem[{{Boyett} {et~al.}(2022){Boyett}, {Stark}, {Bunker}, {Tang}, \&
  {Maseda}}]{boyett_2022a}
{Boyett}, K. N.~K., {Stark}, D.~P., {Bunker}, A.~J., {Tang}, M., \& {Maseda},
  M.~V. 2022, \mnras, 513, 4451, \dodoi{10.1093/mnras/stac1109}

\bibitem[{{Bradley} {et~al.}(2022){Bradley}, {Coe}, {Brammer}, {Furtak},
  {Larson}, {Andrade-Santos}, {Bhatawdekar}, {Bradac}, {Broadhurst}, {Carnall},
  {Conselice}, {Diego}, {Frye}, {Fujimoto}, {Y. -Y Hsiao}, {Hutchison}, {Jung},
  {Mahler}, {McCandliss}, {Oguri}, {Postman}, {Sharon}, {Trenti}, {Vanzella},
  {Welch}, {Windhorst}, \& {Zitrin}}]{Bradley2022highz}
{Bradley}, L.~D., {Coe}, D., {Brammer}, G., {et~al.} 2022, arXiv e-prints,
  arXiv:2210.01777.
\newblock \doarXiv{2210.01777}

\bibitem[{{Brammer} {et~al.}(2022){Brammer}, {Strait}, {Matharu}, \&
  {Momcheva}}]{Grizli}
{Brammer}, G., {Strait}, V., {Matharu}, J., \& {Momcheva}, I. 2022, {grizli},
  1.5.0, Zenodo,  Zenodo, \dodoi{10.5281/zenodo.6672538}

\bibitem[{{Brodie} \& {Strader}(2006)}]{brodie2006}
{Brodie}, J.~P., \& {Strader}, J. 2006, \araa, 44, 193,
  \dodoi{10.1146/annurev.astro.44.051905.092441}

\bibitem[{{Brown} \& {Gnedin}(2021)}]{Brown2021}
{Brown}, G., \& {Gnedin}, O.~Y. 2021, \mnras, 508, 5935,
  \dodoi{10.1093/mnras/stab2907}

\bibitem[{{Bruzual} \& {Charlot}(2003)}]{BC03}
{Bruzual}, G., \& {Charlot}, S. 2003, \mnras, 344, 1000,
  \dodoi{10.1046/j.1365-8711.2003.06897.x}

\bibitem[{{Calzetti} {et~al.}(2000){Calzetti}, {Armus}, {Bohlin}, {Kinney},
  {Koornneef}, \& {Storchi-Bergmann}}]{Calzetti2000}
{Calzetti}, D., {Armus}, L., {Bohlin}, R.~C., {et~al.} 2000, \apj, 533, 682,
  \dodoi{10.1086/308692}

\bibitem[{{Carnall} {et~al.}(2018){Carnall}, {McLure}, {Dunlop}, \&
  {Dav{\'e}}}]{Carnall2018}
{Carnall}, A.~C., {McLure}, R.~J., {Dunlop}, J.~S., \& {Dav{\'e}}, R. 2018,
  \mnras, 480, 4379, \dodoi{10.1093/mnras/sty2169}

\bibitem[{{Chabrier}(2003)}]{Chabrier_2003}
{Chabrier}, G. 2003, \pasp, 115, 763, \dodoi{10.1086/376392}

\bibitem[{{Chevallard} {et~al.}(2018){Chevallard}, {Charlot}, {Senchyna},
  {Stark}, {Vidal-Garc{\'\i}a}, {Feltre}, {Gutkin}, {Jones}, {Mainali}, \&
  {Wofford}}]{chevallard18}
{Chevallard}, J., {Charlot}, S., {Senchyna}, P., {et~al.} 2018, \mnras, 479,
  3264, \dodoi{10.1093/mnras/sty1461}

\bibitem[{{Choi} {et~al.}(2016){Choi}, {Dotter}, {Conroy}, {Cantiello},
  {Paxton}, \& {Johnson}}]{mist_library}
{Choi}, J., {Dotter}, A., {Conroy}, C., {et~al.} 2016, \apj, 823, 102,
  \dodoi{10.3847/0004-637X/823/2/102}

\bibitem[{{Claeyssens} {et~al.}(2022){Claeyssens}, {Adamo}, {Richard},
  {Mahler}, {Messa}, \& {Dessauges-Zavadsky}}]{claeyssens22}
{Claeyssens}, A., {Adamo}, A., {Richard}, J., {et~al.} 2022, arXiv e-prints,
  arXiv:2208.10450.
\newblock \doarXiv{2208.10450}

\bibitem[{{Coe} {et~al.}(2019){Coe}, {Salmon}, {Bradac}, {Bradley}, {Sharon},
  {Zitrin}, {Acebron}, {Cerny}, {Cibirka}, {Strait}, {Paterno-Mahler},
  {Mahler}, {Avila}, {Ogaz}, {Huang}, {Pelliccia}, {Stark}, {Mainali}, {Oesch},
  {Trenti}, {Carrasco}, {Dawson}, {Rodney}, {Strolger}, {Riess}, {Jones},
  {Frye}, {Czakon}, {Umetsu}, {Vulcani}, {Graur}, {Jha}, {Graham}, {Molino},
  {Nonino}, {Hjorth}, {Selsing}, {Christensen}, {Kikuchihara}, {Ouchi},
  {Oguri}, {Welch}, {Lemaux}, {Andrade-Santos}, {Hoag}, {Johnson}, {Peterson},
  {Past}, {Fox}, {Agulli}, {Livermore}, {Ryan}, {Lam}, {Sendra-Server}, {Toft},
  {Lovisari}, \& {Su}}]{Coe_2019}
{Coe}, D., {Salmon}, B., {Bradac}, M., {et~al.} 2019, arXiv e-prints.
\newblock \doarXiv{1903.02002}

\bibitem[{{Conroy} \& {Gunn}(2010)}]{Conroy_2010}
{Conroy}, C., \& {Gunn}, J.~E. 2010, \apj, 712, 833,
  \dodoi{10.1088/0004-637X/712/2/833}

\bibitem[{{Diego} {et~al.}(2005){Diego}, {Protopapas}, {Sandvik}, \&
  {Tegmark}}]{Diego05wslap}
{Diego}, J.~M., {Protopapas}, P., {Sandvik}, H.~B., \& {Tegmark}, M. 2005,
  \mnras, 360, 477, \dodoi{10.1111/j.1365-2966.2005.09021.x}

\bibitem[{{Diego} {et~al.}(2007){Diego}, {Tegmark}, {Protopapas}, \&
  {Sandvik}}]{Diego07wslap2}
{Diego}, J.~M., {Tegmark}, M., {Protopapas}, P., \& {Sandvik}, H.~B. 2007,
  \mnras, 375, 958, \dodoi{10.1111/j.1365-2966.2007.11380.x}

\bibitem[{{Ellis} {et~al.}(2001){Ellis}, {Santos}, {Kneib}, \&
  {Kuijken}}]{ellis2001}
{Ellis}, R., {Santos}, M.~R., {Kneib}, J.-P., \& {Kuijken}, K. 2001, \apjl,
  560, L119, \dodoi{10.1086/324423}

\bibitem[{{Elmegreen} {et~al.}(2020){Elmegreen}, {Elmegreen}, {Whitmore},
  {Chandar}, {Calzetti}, {Lee}, {White}, {Cook}, {Ubeda}, {Mok}, \&
  {Linden}}]{elmegreen20}
{Elmegreen}, D.~M., {Elmegreen}, B.~G., {Whitmore}, B.~C., {et~al.} 2020, arXiv
  e-prints, arXiv:2012.10765.
\newblock \doarXiv{2012.10765}

\bibitem[{{Emami} {et~al.}(2020){Emami}, {Siana}, {Alavi}, {Gburek}, {Freeman},
  {Richard}, {Weisz}, \& {Stark}}]{emami2020_xi}
{Emami}, N., {Siana}, B., {Alavi}, A., {et~al.} 2020, \apj, 895, 116,
  \dodoi{10.3847/1538-4357/ab8f97}

\bibitem[{{Falc{\'o}n-Barroso} {et~al.}(2011){Falc{\'o}n-Barroso},
  {S{\'a}nchez-Bl{\'a}zquez}, {Vazdekis}, {Ricciardelli}, {Cardiel}, {Cenarro},
  {Gorgas}, \& {Peletier}}]{miles_library}
{Falc{\'o}n-Barroso}, J., {S{\'a}nchez-Bl{\'a}zquez}, P., {Vazdekis}, A.,
  {et~al.} 2011, \aap, 532, A95, \dodoi{10.1051/0004-6361/201116842}

\bibitem[{{Ferland} {et~al.}(1998){Ferland}, {Korista}, {Verner}, {Ferguson},
  {Kingdon}, \& {Verner}}]{Ferland1998}
{Ferland}, G.~J., {Korista}, K.~T., {Verner}, D.~A., {et~al.} 1998, \pasp, 110,
  761, \dodoi{10.1086/316190}

\bibitem[{{Ferland} {et~al.}(2013){Ferland}, {Porter}, {van Hoof}, {Williams},
  {Abel}, {Lykins}, {Shaw}, {Henney}, \& {Stancil}}]{Ferland2013}
{Ferland}, G.~J., {Porter}, R.~L., {van Hoof}, P.~A.~M., {et~al.} 2013, \rmxaa,
  49, 137.
\newblock \doarXiv{1302.4485}

\bibitem[{{Ferland} {et~al.}(2017){Ferland}, {Chatzikos}, {Guzm{\'a}n},
  {Lykins}, {van Hoof}, {Williams}, {Abel}, {Badnell}, {Keenan}, {Porter}, \&
  {Stancil}}]{Ferland2017}
{Ferland}, G.~J., {Chatzikos}, M., {Guzm{\'a}n}, F., {et~al.} 2017, \rmxaa, 53,
  385.
\newblock \doarXiv{1705.10877}

\bibitem[{{Forbes} \& {Bridges}(2010)}]{forbes2010}
{Forbes}, D.~A., \& {Bridges}, T. 2010, \mnras, 404, 1203,
  \dodoi{10.1111/j.1365-2966.2010.16373.x}

\bibitem[{{Gaia Collaboration} {et~al.}(2021){Gaia Collaboration}, {Brown},
  {Vallenari}, {Prusti}, {de Bruijne}, {Babusiaux}, {Biermann}, {Creevey},
  {Evans}, {Eyer}, {Hutton}, {Jansen}, {Jordi}, {Klioner}, {Lammers},
  {Lindegren}, {Luri}, {Mignard}, {Panem}, {Pourbaix}, {Randich}, {Sartoretti},
  {Soubiran}, {Walton}, {Arenou}, {Bailer-Jones}, {Bastian}, {Cropper},
  {Drimmel}, {Katz}, {Lattanzi}, {van Leeuwen}, {Bakker}, {Cacciari},
  {Casta{\~n}eda}, {De Angeli}, {Ducourant}, {Fabricius}, {Fouesneau},
  {Fr{\'e}mat}, {Guerra}, {Guerrier}, {Guiraud}, {Jean-Antoine Piccolo},
  {Masana}, {Messineo}, {Mowlavi}, {Nicolas}, {Nienartowicz}, {Pailler},
  {Panuzzo}, {Riclet}, {Roux}, {Seabroke}, {Sordo}, {Tanga}, {Th{\'e}venin},
  {Gracia-Abril}, {Portell}, {Teyssier}, {Altmann}, {Andrae}, {Bellas-Velidis},
  {Benson}, {Berthier}, {Blomme}, {Brugaletta}, {Burgess}, {Busso}, {Carry},
  {Cellino}, {Cheek}, {Clementini}, {Damerdji}, {Davidson}, {Delchambre},
  {Dell'Oro}, {Fern{\'a}ndez-Hern{\'a}ndez}, {Galluccio}, {Garc{\'\i}a-Lario},
  {Garcia-Reinaldos}, {Gonz{\'a}lez-N{\'u}{\~n}ez}, {Gosset}, {Haigron},
  {Halbwachs}, {Hambly}, {Harrison}, {Hatzidimitriou}, {Heiter},
  {Hern{\'a}ndez}, {Hestroffer}, {Hodgkin}, {Holl}, {Jan{\ss}en}, {Jevardat de
  Fombelle}, {Jordan}, {Krone-Martins}, {Lanzafame}, {L{\"o}ffler}, {Lorca},
  {Manteiga}, {Marchal}, {Marrese}, {Moitinho}, {Mora}, {Muinonen}, {Osborne},
  {Pancino}, {Pauwels}, {Petit}, {Recio-Blanco}, {Richards}, {Riello},
  {Rimoldini}, {Robin}, {Roegiers}, {Rybizki}, {Sarro}, {Siopis}, {Smith},
  {Sozzetti}, {Ulla}, {Utrilla}, {van Leeuwen}, {van Reeven}, {Abbas}, {Abreu
  Aramburu}, {Accart}, {Aerts}, {Aguado}, {Ajaj}, {Altavilla}, {{\'A}lvarez},
  {{\'A}lvarez Cid-Fuentes}, {Alves}, {Anderson}, {Anglada Varela}, {Antoja},
  {Audard}, {Baines}, {Baker}, {Balaguer-N{\'u}{\~n}ez}, {Balbinot}, {Balog},
  {Barache}, {Barbato}, {Barros}, {Barstow}, {Bartolom{\'e}}, {Bassilana},
  {Bauchet}, {Baudesson-Stella}, {Becciani}, {Bellazzini}, {Bernet}, {Bertone},
  {Bianchi}, {Blanco-Cuaresma}, {Boch}, {Bombrun}, {Bossini}, {Bouquillon},
  {Bragaglia}, {Bramante}, {Breedt}, {Bressan}, {Brouillet}, {Bucciarelli},
  {Burlacu}, {Busonero}, {Butkevich}, {Buzzi}, {Caffau}, {Cancelliere},
  {C{\'a}novas}, {Cantat-Gaudin}, {Carballo}, {Carlucci}, {Carnerero},
  {Carrasco}, {Casamiquela}, {Castellani}, {Castro-Ginard}, {Castro Sampol},
  {Chaoul}, {Charlot}, {Chemin}, {Chiavassa}, {Cioni}, {Comoretto}, {Cooper},
  {Cornez}, {Cowell}, {Crifo}, {Crosta}, {Crowley}, {Dafonte}, {Dapergolas},
  {David}, {David}, {de Laverny}, {De Luise}, {De March}, {De Ridder}, {de
  Souza}, {de Teodoro}, {de Torres}, {del Peloso}, {del Pozo}, {Delbo},
  {Delgado}, {Delgado}, {Delisle}, {Di Matteo}, {Diakite}, {Diener},
  {Distefano}, {Dolding}, {Eappachen}, {Edvardsson}, {Enke}, {Esquej}, {Fabre},
  {Fabrizio}, {Faigler}, {Fedorets}, {Fernique}, {Fienga}, {Figueras},
  {Fouron}, {Fragkoudi}, {Fraile}, {Franke}, {Gai}, {Garabato},
  {Garcia-Gutierrez}, {Garc{\'\i}a-Torres}, {Garofalo}, {Gavras}, {Gerlach},
  {Geyer}, {Giacobbe}, {Gilmore}, {Girona}, {Giuffrida}, {Gomel}, {Gomez},
  {Gonzalez-Santamaria}, {Gonz{\'a}lez-Vidal}, {Granvik},
  {Guti{\'e}rrez-S{\'a}nchez}, {Guy}, {Hauser}, {Haywood}, {Helmi}, {Hidalgo},
  {Hilger}, {H{\l}adczuk}, {Hobbs}, {Holland}, {Huckle}, {Jasniewicz},
  {Jonker}, {Juaristi Campillo}, {Julbe}, {Karbevska}, {Kervella}, {Khanna},
  {Kochoska}, {Kontizas}, {Kordopatis}, {Korn}, {Kostrzewa-Rutkowska},
  {Kruszy{\'n}ska}, {Lambert}, {Lanza}, {Lasne}, {Le Campion}, {Le Fustec},
  {Lebreton}, {Lebzelter}, {Leccia}, {Leclerc}, {Lecoeur-Taibi}, {Liao},
  {Licata}, {Lindstr{\o}m}, {Lister}, {Livanou}, {Lobel}, {Madrero Pardo},
  {Managau}, {Mann}, {Marchant}, {Marconi}, {Marcos Santos}, {Marinoni},
  {Marocco}, {Marshall}, {Martin Polo}, {Mart{\'\i}n-Fleitas}, {Masip},
  {Massari}, {Mastrobuono-Battisti}, {Mazeh}, {McMillan}, {Messina},
  {Michalik}, {Millar}, {Mints}, {Molina}, {Molinaro}, {Moln{\'a}r},
  {Montegriffo}, {Mor}, {Morbidelli}, {Morel}, {Morris}, {Mulone}, {Munoz},
  {Muraveva}, {Murphy}, {Musella}, {Noval}, {Ord{\'e}novic}, {Orr{\`u}},
  {Osinde}, {Pagani}, {Pagano}, {Palaversa}, {Palicio}, {Panahi}, {Pawlak},
  {Pe{\~n}alosa Esteller}, {Penttil{\"a}}, {Piersimoni}, {Pineau}, {Plachy},
  {Plum}, {Poggio}, {Poretti}, {Poujoulet}, {Pr{\v{s}}a}, {Pulone}, {Racero},
  {Ragaini}, {Rainer}, {Raiteri}, {Rambaux}, {Ramos}, {Ramos-Lerate}, {Re
  Fiorentin}, {Regibo}, {Reyl{\'e}}, {Ripepi}, {Riva}, {Rixon}, {Robichon},
  {Robin}, {Roelens}, {Rohrbasser}, {Romero-G{\'o}mez}, {Rowell}, {Royer},
  {Rybicki}, {Sadowski}, {Sagrist{\`a} Sell{\'e}s}, {Sahlmann}, {Salgado},
  {Salguero}, {Samaras}, {Sanchez Gimenez}, {Sanna}, {Santove{\~n}a},
  {Sarasso}, {Schultheis}, {Sciacca}, {Segol}, {Segovia}, {S{\'e}gransan},
  {Semeux}, {Shahaf}, {Siddiqui}, {Siebert}, {Siltala}, {Slezak}, {Smart},
  {Solano}, {Solitro}, {Souami}, {Souchay}, {Spagna}, {Spoto}, {Steele},
  {Steidelm{\"u}ller}, {Stephenson}, {S{\"u}veges}, {Szabados}, {Szegedi-Elek},
  {Taris}, {Tauran}, {Taylor}, {Teixeira}, {Thuillot}, {Tonello}, {Torra},
  {Torra}, {Turon}, {Unger}, {Vaillant}, {van Dillen}, {Vanel}, {Vecchiato},
  {Viala}, {Vicente}, {Voutsinas}, {Weiler}, {Wevers}, {Wyrzykowski}, {Yoldas},
  {Yvard}, {Zhao}, {Zorec}, {Zucker}, {Zurbach}, \& {Zwitter}}]{Gaia_EDR3}
{Gaia Collaboration}, {Brown}, A.~G.~A., {Vallenari}, A., {et~al.} 2021, \aap,
  649, A1, \dodoi{10.1051/0004-6361/202039657}

\bibitem[{{Gieles} \& {Portegies Zwart}(2011)}]{gieles11}
{Gieles}, M., \& {Portegies Zwart}, S.~F. 2011, \mnras, 410, L6,
  \dodoi{10.1111/j.1745-3933.2010.00967.x}

\bibitem[{{G{\"o}tberg} {et~al.}(2018){G{\"o}tberg}, {de Mink}, {Groh},
  {Kupfer}, {Crowther}, {Zapartas}, \& {Renzo}}]{Goetberg2018}
{G{\"o}tberg}, Y., {de Mink}, S.~E., {Groh}, J.~H., {et~al.} 2018, \aap, 615,
  A78, \dodoi{10.1051/0004-6361/201732274}

\bibitem[{{G{\"o}tberg} {et~al.}(2019){G{\"o}tberg}, {de Mink}, {Groh},
  {Leitherer}, \& {Norman}}]{Goetberg2019}
{G{\"o}tberg}, Y., {de Mink}, S.~E., {Groh}, J.~H., {Leitherer}, C., \&
  {Norman}, C. 2019, \aap, 629, A134, \dodoi{10.1051/0004-6361/201834525}

\bibitem[{{Heckman} {et~al.}(2011){Heckman}, {Borthakur}, {Overzier},
  {Kauffmann}, {Basu-Zych}, {Leitherer}, {Sembach}, {Martin}, {Rich},
  {Schiminovich}, \& {Seibert}}]{heckman11}
{Heckman}, T.~M., {Borthakur}, S., {Overzier}, R., {et~al.} 2011, \apj, 730, 5,
  \dodoi{10.1088/0004-637X/730/1/5}

\bibitem[{{Herenz} {et~al.}(2017){Herenz}, {Hayes}, {Papaderos}, {Cannon},
  {Bik}, {Melinder}, \& {{\"O}stlin}}]{herenz17_ionized_channel}
{Herenz}, E.~C., {Hayes}, M., {Papaderos}, P., {et~al.} 2017, \aap, 606, L11,
  \dodoi{10.1051/0004-6361/201731809}

\bibitem[{{Hsiao} {et~al.}(2022){Hsiao}, {Coe}, {Abdurrouf}, {Whitler}, {Jung},
  {Khullar}, {Meena}, {Dayal}, {Barrow}, {Santos-Olmsted}, {Casselman},
  {Vanzella}, {Nonino}, {Jimenez-Teja}, {Oguri}, {Stark}, {Furtak}, {Zitrin},
  {Adamo}, {Brammer}, {Bradley}, {Diego}, {Zackrisson}, {Finkelstein},
  {Windhorst}, {Bhatawdekar}, {Hutchison}, {Broadhurst}, {Dimauro},
  {Andrade-Santos}, {Eldridge}, {Acebron}, {Avila}, {Bayliss}, {Benitez},
  {Binggeli}, {Bolan}, {Bradac}, {Carnall}, {Conselice}, {Donahue}, {Frye},
  {Fujimoto}, {Henry}, {James}, {Kassin}, {Kewley}, {Larson}, {Lauer}, {Law},
  {Mahler}, {Mainali}, {McCandliss}, {Nicholls}, {Pirzkal}, {Postman}, {Rigby},
  {Ryan}, {Senchyna}, {Sharon}, {Shimizu}, {Strait}, {Tang}, {Trenti},
  {Vikaeus}, \& {Welch}}]{Hsiao2022}
{Hsiao}, T., {Coe}, D., {Abdurrouf}, {et~al.} 2022, arXiv e-prints,
  arXiv:2210.14123.
\newblock \doarXiv{2210.14123}

\bibitem[{{James} {et~al.}(2016){James}, {Auger}, {Aloisi}, {Calzetti}, \&
  {Kewley}}]{james16}
{James}, B.~L., {Auger}, M., {Aloisi}, A., {Calzetti}, D., \& {Kewley}, L.
  2016, \apj, 816, 40, \dodoi{10.3847/0004-637X/816/1/40}

\bibitem[{{Johnson} {et~al.}(2021){Johnson}, {Leja}, {Conroy}, \&
  {Speagle}}]{prospector}
{Johnson}, B.~D., {Leja}, J., {Conroy}, C., \& {Speagle}, J.~S. 2021, \apjs,
  254, 22, \dodoi{10.3847/1538-4365/abef67}

\bibitem[{{Johnson} {et~al.}(2017){Johnson}, {Rigby}, {Sharon}, {Gladders},
  {Florian}, {Bayliss}, {Wuyts}, {Whitaker}, {Livermore}, \&
  {Murray}}]{johnson17}
{Johnson}, T.~L., {Rigby}, J.~R., {Sharon}, K., {et~al.} 2017, \apjl, 843, L21,
  \dodoi{10.3847/2041-8213/aa7516}

\bibitem[{{Jullo} \& {Kneib}(2009)}]{Jullo_Kneib_lenstool}
{Jullo}, E., \& {Kneib}, J.-P. 2009, \mnras, 395, 1319,
  \dodoi{10.1111/j.1365-2966.2009.14654.x}

\bibitem[{{Jullo} {et~al.}(2007){Jullo}, {Kneib}, {Limousin},
  {El{\'{\i}}asd{\'o}ttir}, {Marshall}, \& {Verdugo}}]{Jullo_lenstool}
{Jullo}, E., {Kneib}, J.-P., {Limousin}, M., {et~al.} 2007, New Journal of
  Physics, 9, 447, \dodoi{10.1088/1367-2630/9/12/447}

\bibitem[{{Khullar} {et~al.}(2021){Khullar}, {Gozman}, {Lin}, {Martinez},
  {Matthews Acu{\~n}a}, {Medina}, {Merz}, {Sanchez}, {Sisco}, {Kavin Stein},
  {Sukay}, {Tavangar}, {Bayliss}, {Bleem}, {Brownsberger}, {Dahle}, {Florian},
  {Gladders}, {Mahler}, {Rigby}, {Sharon}, \& {Stark}}]{Khullar2021}
{Khullar}, G., {Gozman}, K., {Lin}, J.~J., {et~al.} 2021, \apj, 906, 107,
  \dodoi{10.3847/1538-4357/abcb86}

\bibitem[{{Kokorev} {et~al.}(2022){Kokorev}, {Brammer}, {Fujimoto}, {Kohno},
  {Magdis}, {Valentino}, {Toft}, {Oesch}, {Bauer}, {Coe}, {Egami}, {Oguri},
  {Ouchi}, {Postman}, {Richard}, {Jolly}, {Knudsen}, {Sun}, {Weaver}, {Ao},
  {Baker}, {Caputi}, {Espada}, {Hatsukade}, {Koekemoer}, {Mu{\~n}oz Arancibia},
  {Shimasaku}, {Umehata}, {Wang}, \& {Wang}}]{kokorev22}
{Kokorev}, V., {Brammer}, G., {Fujimoto}, S., {et~al.} 2022, arXiv e-prints,
  arXiv:2207.07125.
\newblock \doarXiv{2207.07125}

\bibitem[{{Kriek} \& {Conroy}(2013)}]{Kriek_2013}
{Kriek}, M., \& {Conroy}, C. 2013, \apjl, 775, L16,
  \dodoi{10.1088/2041-8205/775/1/L16}

\bibitem[{{Kroupa}(2001)}]{Kroupa2001}
{Kroupa}, P. 2001, \mnras, 322, 231, \dodoi{10.1046/j.1365-8711.2001.04022.x}

\bibitem[{{Kroupa}(2002)}]{Kroupa2002}
---. 2002, Science, 295, 82, \dodoi{10.1126/science.1067524}

\bibitem[{{Krumholz} {et~al.}(2019){Krumholz}, {McKee}, \&
  {Bland-Hawthorn}}]{krumholz2019}
{Krumholz}, M.~R., {McKee}, C.~F., \& {Bland-Hawthorn}, J. 2019, \araa, 57,
  227, \dodoi{10.1146/annurev-astro-091918-104430}

\bibitem[{{Larsen} {et~al.}(2012){Larsen}, {Strader}, \& {Brodie}}]{larsen2012}
{Larsen}, S.~S., {Strader}, J., \& {Brodie}, J.~P. 2012, \aap, 544, L14,
  \dodoi{10.1051/0004-6361/201219897}

\bibitem[{{Leitherer} {et~al.}(2014){Leitherer}, {Ekstr{\"o}m}, {Meynet},
  {Schaerer}, {Agienko}, \& {Levesque}}]{leitherer14}
{Leitherer}, C., {Ekstr{\"o}m}, S., {Meynet}, G., {et~al.} 2014, \apjs, 212,
  14, \dodoi{10.1088/0067-0049/212/1/14}

\bibitem[{{Leja} {et~al.}(2019){Leja}, {Carnall}, {Johnson}, {Conroy}, \&
  {Speagle}}]{Leja_2019}
{Leja}, J., {Carnall}, A.~C., {Johnson}, B.~D., {Conroy}, C., \& {Speagle},
  J.~S. 2019, \apj, 876, 3, \dodoi{10.3847/1538-4357/ab133c}

\bibitem[{{Livermore} {et~al.}(2017){Livermore}, {Finkelstein}, \&
  {Lotz}}]{livermore17}
{Livermore}, R.~C., {Finkelstein}, S.~L., \& {Lotz}, J.~M. 2017, \apj, 835,
  113, \dodoi{10.3847/1538-4357/835/2/113}

\bibitem[{{Mainali} {et~al.}(2022){Mainali}, {Rigby}, {Chisholm}, {Bayliss},
  {Bordoloi}, {Gladders}, {Rivera-Thorsen}, {Dahle}, {Sharon}, {Florian},
  {Berg}, {Sharma}, {Riley Owens}, {Kjellgren}, {Kim}, \& {Wayne}}]{mainali22}
{Mainali}, R., {Rigby}, J.~R., {Chisholm}, J., {et~al.} 2022, arXiv e-prints,
  arXiv:2210.11575.
\newblock \doarXiv{2210.11575}

\bibitem[{{Maseda} {et~al.}(2020){Maseda}, {Bacon}, {Lam}, {Matthee},
  {Brinchmann}, {Schaye}, {Labbe}, {Schmidt}, {Boogaard}, {Bouwens},
  {Cantalupo}, {Franx}, {Hashimoto}, {Inami}, {Kusakabe}, {Mahler},
  {Nanayakkara}, {Richard}, \& {Wisotzki}}]{maseda20}
{Maseda}, M.~V., {Bacon}, R., {Lam}, D., {et~al.} 2020, \mnras, 493, 5120,
  \dodoi{10.1093/mnras/staa622}

\bibitem[{Matthee {et~al.}(2022)Matthee, Mackenzie, Simcoe, Kashino, Lilly,
  Bordoloi, \& Eilers}]{matthee22_nircam_slitless}
Matthee, J., Mackenzie, R., Simcoe, R.~A., {et~al.} 2022

\bibitem[{{Messa} {et~al.}(2022){Messa}, {Dessauges-Zavadsky}, {Richard},
  {Adamo}, {Nagy}, {Combes}, {Mayer}, \& {Ebeling}}]{messa2022}
{Messa}, M., {Dessauges-Zavadsky}, M., {Richard}, J., {et~al.} 2022, \mnras,
  516, 2420, \dodoi{10.1093/mnras/stac2189}

\bibitem[{{Me{\v{s}}tri{\'c}} {et~al.}(2022){Me{\v{s}}tri{\'c}}, {Vanzella},
  {Zanella}, {Castellano}, {Calura}, {Rosati}, {Bergamini}, {Mercurio},
  {Meneghetti}, {Grillo}, {Caminha}, {Nonino}, {Merlin}, {Cupani}, \&
  {Sani}}]{Mestric22}
{Me{\v{s}}tri{\'c}}, U., {Vanzella}, E., {Zanella}, A., {et~al.} 2022, arXiv
  e-prints, arXiv:2202.09377, \dodoi{10.48550/arXiv.2202.09377}

\bibitem[{{Micheva} {et~al.}(2017){Micheva}, {Oey}, {Jaskot}, \&
  {James}}]{micheva17}
{Micheva}, G., {Oey}, M.~S., {Jaskot}, A.~E., \& {James}, B.~L. 2017, \apj,
  845, 165, \dodoi{10.3847/1538-4357/aa830b}

\bibitem[{{Mowla} {et~al.}(2022){Mowla}, {Iyer}, {Desprez},
  {Estrada-Carpenter}, {Martis}, {Noirot}, {Sarrouh}, {Strait}, {Asada},
  {Abraham}, {Brammer}, {Sawicki}, {Willott}, {Bradac}, {Doyon}, {Muzzin},
  {Pacifici}, {Ravindranath}, \& {Zabl}}]{mowla22}
{Mowla}, L., {Iyer}, K.~G., {Desprez}, G., {et~al.} 2022, \apjl, 937, L35,
  \dodoi{10.3847/2041-8213/ac90ca}

\bibitem[{{Nakajima} {et~al.}(2020){Nakajima}, {Ellis}, {Robertson}, {Tang}, \&
  {Stark}}]{nakajima2020}
{Nakajima}, K., {Ellis}, R.~S., {Robertson}, B.~E., {Tang}, M., \& {Stark},
  D.~P. 2020, \apj, 889, 161, \dodoi{10.3847/1538-4357/ab6604}

\bibitem[{{Nakajima} {et~al.}(2022){Nakajima}, {Ouchi}, {Xu}, {Rauch},
  {Harikane}, {Nishigaki}, {Isobe}, {Kusakabe}, {Nagao}, {Ono}, {Onodera},
  {Sugahara}, {Kim}, {Komiyama}, {Lee}, \& {Zahedy}}]{nakajima2022}
{Nakajima}, K., {Ouchi}, M., {Xu}, Y., {et~al.} 2022, \apjs, 262, 3,
  \dodoi{10.3847/1538-4365/ac7710}

\bibitem[{{Narloch} {et~al.}(2022){Narloch}, {Pietrzy{\'n}ski}, {Gieren},
  {Piatti}, {Karczmarek}, {G{\'o}rski}, {Graczyk}, {Smolec}, {Hajdu},
  {Suchomska}, {Zgirski}, {Wielg{\'o}rski}, {Pilecki}, {Taormina},
  {Ka{\l}uszy{\'n}ski}, {Pych}, {Rojas Garc{\'\i}a}, \& {Lewis}}]{narloch2022}
{Narloch}, W., {Pietrzy{\'n}ski}, G., {Gieren}, W., {et~al.} 2022, \aap, 666,
  A80, \dodoi{10.1051/0004-6361/202243378}

\bibitem[{{Norris} {et~al.}(2014){Norris}, {Kannappan}, {Forbes}, {Romanowsky},
  {Brodie}, {Faifer}, {Huxor}, {Maraston}, {Moffett}, {Penny}, {Pota},
  {Smith-Castelli}, {Strader}, {Bradley}, {Eckert}, {Fohring}, {McBride},
  {Stark}, \& {Vaduvescu}}]{norris2014}
{Norris}, M.~A., {Kannappan}, S.~J., {Forbes}, D.~A., {et~al.} 2014, \mnras,
  443, 1151, \dodoi{10.1093/mnras/stu1186}

\bibitem[{{Oguri}(2010)}]{Oguri_2010}
{Oguri}, M. 2010, \pasj, 62, 1017, \dodoi{10.1093/pasj/62.4.1017}

\bibitem[{{Oguri}(2021)}]{Oguri_2021}
---. 2021, \pasp, 133, 074504, \dodoi{10.1088/1538-3873/ac12db}

\bibitem[{{Oke} \& {Gunn}(1983)}]{Oke_1983}
{Oke}, J.~B., \& {Gunn}, J.~E. 1983, The Astrophysical Journal, 266, 713,
  \dodoi{10.1086/160817}

\bibitem[{{{\"O}stlin} {et~al.}(2007){{\"O}stlin}, {Cumming}, \&
  {Bergvall}}]{ostlin2007}
{{\"O}stlin}, G., {Cumming}, R.~J., \& {Bergvall}, N. 2007, \aap, 461, 471,
  \dodoi{10.1051/0004-6361:20054461}

\bibitem[{{Plazas} {et~al.}(2019){Plazas}, {Meneghetti}, {Maturi}, \&
  {Rhodes}}]{Plazas_2019}
{Plazas}, A.~A., {Meneghetti}, M., {Maturi}, M., \& {Rhodes}, J. 2019, \mnras,
  482, 2823, \dodoi{10.1093/mnras/sty2737}

\bibitem[{{Pontoppidan} {et~al.}(2022){Pontoppidan}, {Blome}, {Braun}, {Brown},
  {Carruthers}, {Coe}, {DePasquale}, {Espinoza}, {Garcia Marin}, {Gordon},
  {Henry}, {Hustak}, {James}, {Koekemoer}, {LaMassa}, {Law}, {Lockwood},
  {Moro-Martin}, {Mullally}, {Pagan}, {Player}, {Proffitt}, {Pulliam},
  {Ramsay}, {Ravindranath}, {Reid}, {Robberto}, {Sabbi}, \&
  {Ubeda}}]{Pontoppidan2022}
{Pontoppidan}, K., {Blome}, C., {Braun}, H., {et~al.} 2022, arXiv e-prints,
  arXiv:2207.13067.
\newblock \doarXiv{2207.13067}

\bibitem[{{Recio-Blanco}(2018)}]{recioblanco2018}
{Recio-Blanco}, A. 2018, \aap, 620, A194, \dodoi{10.1051/0004-6361/201833179}

\bibitem[{{Reina-Campos} {et~al.}(2018){Reina-Campos}, {Kruijssen}, {Pfeffer},
  {Bastian}, \& {Crain}}]{reinacampos2018}
{Reina-Campos}, M., {Kruijssen}, J.~M.~D., {Pfeffer}, J., {Bastian}, N., \&
  {Crain}, R.~A. 2018, \mnras, 481, 2851, \dodoi{10.1093/mnras/sty2451}

\bibitem[{{Rigby} {et~al.}(2017){Rigby}, {Johnson}, {Sharon}, {Whitaker},
  {Gladders}, {Florian}, {Lotz}, {Bayliss}, \& {Wuyts}}]{rigby17}
{Rigby}, J.~R., {Johnson}, T.~L., {Sharon}, K., {et~al.} 2017, \apj, 843, 79,
  \dodoi{10.3847/1538-4357/aa775e}

\bibitem[{{Rivera-Thorsen} {et~al.}(2019){Rivera-Thorsen}, {Dahle}, {Chisholm},
  {Florian}, {Gronke}, {Rigby}, {Gladders}, {Mahler}, {Sharon}, \&
  {Bayliss}}]{rivera19}
{Rivera-Thorsen}, T.~E., {Dahle}, H., {Chisholm}, J., {et~al.} 2019, Science,
  366, 738, \dodoi{10.1126/science.aaw0978}

\bibitem[{{Ryon} {et~al.}(2017){Ryon}, {Gallagher}, {Smith}, {Adamo},
  {Calzetti}, {Bright}, {Cignoni}, {Cook}, {Dale}, {Elmegreen}, {Fumagalli},
  {Gouliermis}, {Grasha}, {Grebel}, {Kim}, {Messa}, {Thilker}, \&
  {Ubeda}}]{ryon17}
{Ryon}, J.~E., {Gallagher}, J.~S., {Smith}, L.~J., {et~al.} 2017, \apj, 841,
  92, \dodoi{10.3847/1538-4357/aa719e}

\bibitem[{{Salim} {et~al.}(2018){Salim}, {Boquien}, \& {Lee}}]{Salim18}
{Salim}, S., {Boquien}, M., \& {Lee}, J.~C. 2018, \apj, 859, 11,
  \dodoi{10.3847/1538-4357/aabf3c}

\bibitem[{{Salmon} {et~al.}(2020){Salmon}, {Coe}, {Bradley}, {Bouwens},
  {Brada{\v{c}}}, {Huang}, {Oesch}, {Stark}, {Sharon}, {Trenti}, {Avila},
  {Ogaz}, {Andrade-Santos}, {Carrasco}, {Cerny}, {Dawson}, {Frye}, {Hoag},
  {Johnson}, {Jones}, {Lam}, {Lovisari}, {Mainali}, {Past}, {Paterno-Mahler},
  {Peterson}, {Riess}, {Rodney}, {Ryan}, {Sendra-Server}, {Strait}, {Strolger},
  {Umetsu}, {Vulcani}, \& {Zitrin}}]{Salmon2020}
{Salmon}, B., {Coe}, D., {Bradley}, L., {et~al.} 2020, \apj, 889, 189,
  \dodoi{10.3847/1538-4357/ab5a8b}

\bibitem[{{Sharon} {et~al.}(2022){Sharon}, {Mahler}, {Rivera-Thorsen}, {Dahle},
  {Gladders}, {Bayliss}, {Florian}, {Kim}, {Khullar}, {Mainali}, {Napier},
  {Navarre}, {Rigby}, {Remolina Gonzalez}, \& {Sharma}}]{sharon2022}
{Sharon}, K., {Mahler}, G., {Rivera-Thorsen}, T.~E., {et~al.} 2022, arXiv
  e-prints, arXiv:2209.03417.
\newblock \doarXiv{2209.03417}

\bibitem[{{Sirressi} {et~al.}(2022){Sirressi}, {Adamo}, {Hayes}, {Bik},
  {Strand{\"a}nger}, {Runnholm}, {Oey}, {{\"O}stlin}, {Menacho}, \&
  {Smith}}]{sirressi22_HARO11}
{Sirressi}, M., {Adamo}, A., {Hayes}, M., {et~al.} 2022, \mnras, 510, 4819,
  \dodoi{10.1093/mnras/stab3774}

\bibitem[{{Stanway} \& {Eldridge}(2018)}]{BPASS}
{Stanway}, E.~R., \& {Eldridge}, J.~J. 2018, \mnras, 479, 75,
  \dodoi{10.1093/mnras/sty1353}

\bibitem[{{Sukay} {et~al.}(2022){Sukay}, {Khullar}, {Gladders}, {Sharon},
  {Mahler}, {Napier}, {Bleem}, {Dahle}, {Florian}, {Gozman}, {Lin}, {Martinez},
  {Matthews Acu{\~n}a}, {Medina}, {Merz}, {Sanchez}, {Sisco}, {Kavin Stein},
  {Tavangar}, \& {Whitaker}}]{Sukay2022}
{Sukay}, E., {Khullar}, G., {Gladders}, M.~D., {et~al.} 2022, arXiv e-prints,
  arXiv:2203.11957.
\newblock \doarXiv{2203.11957}

\bibitem[{{Trebitsch} {et~al.}(2017){Trebitsch}, {Blaizot}, {Rosdahl},
  {Devriendt}, \& {Slyz}}]{trebitsch2017}
{Trebitsch}, M., {Blaizot}, J., {Rosdahl}, J., {Devriendt}, J., \& {Slyz}, A.
  2017, \mnras, 470, 224, \dodoi{10.1093/mnras/stx1060}

\bibitem[{{Treu} {et~al.}(2022){Treu}, {Roberts-Borsani}, {Bradac}, {Brammer},
  {Fontana}, {Henry}, {Mason}, {Morishita}, {Pentericci}, {Wang}, {Acebron},
  {Bagley}, {Bergamini}, {Belfiori}, {Bonchi}, {Boyett}, {Boutsia}, {Calabro},
  {Caminha}, {Castellano}, {Dressler}, {Glazebrook}, {Grillo}, {Jacobs},
  {Jones}, {Kelly}, {Leethochawalit}, {Malkan}, {Marchesini}, {Mascia},
  {Mercurio}, {Merlin}, {Nanayakkara}, {Nonino}, {Paris}, {Poggianti},
  {Rosati}, {Santini}, {Scarlata}, {Shipley}, {Strait}, {Trenti}, {Tubthong},
  {Vanzella}, {Vulcani}, \& {Yang}}]{TreuGlass22}
{Treu}, T., {Roberts-Borsani}, G., {Bradac}, M., {et~al.} 2022, ApJ, in press,
  arXiv:2206.07978.
\newblock \doarXiv{2206.07978}

\bibitem[{{Vanzella} {et~al.}(2017){Vanzella}, {Calura}, {Meneghetti},
  {Mercurio}, {Castellano}, {Caminha}, {Balestra}, {Rosati}, {Tozzi}, {De
  Barros}, {Grazian}, {D'Ercole}, {Ciotti}, {Caputi}, {Grillo}, {Merlin},
  {Pentericci}, {Fontana}, {Cristiani}, \& {Coe}}]{vanz_paving}
{Vanzella}, E., {Calura}, F., {Meneghetti}, M., {et~al.} 2017, \mnras, 467,
  4304, \dodoi{10.1093/mnras/stx351}

\bibitem[{{Vanzella} {et~al.}(2019){Vanzella}, {Calura}, {Meneghetti},
  {Castellano}, {Caminha}, {Mercurio}, {Cupani}, {Rosati}, {Grillo}, {Gilli},
  {Mignoli}, {Fiorentino}, {Arcidiacono}, {Lombini}, \& {Cortecchia}}]{vanz19}
---. 2019, \mnras, 483, 3618, \dodoi{10.1093/mnras/sty3311}

\bibitem[{{Vanzella} {et~al.}(2021){Vanzella}, {Caminha}, {Rosati}, {Mercurio},
  {Castellano}, {Meneghetti}, {Grillo}, {Sani}, {Bergamini}, {Calura},
  {Caputi}, {Cristiani}, {Cupani}, {Fontana}, {Gilli}, {Grazian}, {Gronke},
  {Mignoli}, {Nonino}, {Pentericci}, {Tozzi}, {Treu}, {Balestra}, \&
  {Dijkstra}}]{vanz_mdlf}
{Vanzella}, E., {Caminha}, G.~B., {Rosati}, P., {et~al.} 2021, \aap, 646, A57,
  \dodoi{10.1051/0004-6361/202039466}

\bibitem[{{Vanzella} {et~al.}(2022{\natexlab{a}}){Vanzella}, {Castellano},
  {Bergamini}, {Meneghetti}, {Zanella}, {Calura}, {Caminha}, {Rosati},
  {Cupani}, {Me{\v{s}}tri{\'c}}, {Brammer}, {Tozzi}, {Mercurio}, {Grillo},
  {Sani}, {Cristiani}, {Nonino}, {Merlin}, \& {Pignataro}}]{vanz22_CFE}
{Vanzella}, E., {Castellano}, M., {Bergamini}, P., {et~al.} 2022{\natexlab{a}},
  \aap, 659, A2, \dodoi{10.1051/0004-6361/202141590}

\bibitem[{{Vanzella} {et~al.}(2022{\natexlab{b}}){Vanzella}, {Castellano},
  {Bergamini}, {Treu}, {Mercurio}, {Scarlata}, {Rosati}, {Grillo}, {Acebron},
  {Caminha}, {Nonino}, {Nanayakkara}, {Roberts-Borsani}, {Bradac}, {Wang},
  {Brammer}, {Strait}, {Vulcani}, {Mestric}, {Meneghetti}, {Calura}, {Henry},
  {Zanella}, {Trenti}, {Boyett}, {Morishita}, {Calabro}, {Glazebrook},
  {Marchesini}, {Birrer}, {Yang}, \& {Jones}}]{vanz_glass22}
---. 2022{\natexlab{b}}, arXiv e-prints, arXiv:2208.00520.
\newblock \doarXiv{2208.00520}

\bibitem[{{Vazdekis} {et~al.}(2016){Vazdekis}, {Koleva}, {Ricciardelli},
  {R{\"o}ck}, \& {Falc{\'o}n-Barroso}}]{miles_2016}
{Vazdekis}, A., {Koleva}, M., {Ricciardelli}, E., {R{\"o}ck}, B., \&
  {Falc{\'o}n-Barroso}, J. 2016, \mnras, 463, 3409,
  \dodoi{10.1093/mnras/stw2231}

\bibitem[{{Vazdekis} {et~al.}(2015){Vazdekis}, {Coelho}, {Cassisi},
  {Ricciardelli}, {Falc{\'o}n-Barroso}, {S{\'a}nchez-Bl{\'a}zquez}, {La
  Barbera}, {Beasley}, \& {Pietrinferni}}]{vazdekis2015}
{Vazdekis}, A., {Coelho}, P., {Cassisi}, S., {et~al.} 2015, \mnras, 449, 1177,
  \dodoi{10.1093/mnras/stv151}

\bibitem[{{Welch} {et~al.}(2022{\natexlab{a}}){Welch}, {Coe}, {Zitrin},
  {Diego}, {Windhorst}, {Mandelker}, {Vanzella}, {Ravindranath}, {Zackrisson},
  {Florian}, {Bradley}, {Sharon}, {Brada{\v{c}}}, {Rigby}, {Frye}, \&
  {Fujimoto}}]{Welch22_clumps}
{Welch}, B., {Coe}, D., {Zitrin}, A., {et~al.} 2022{\natexlab{a}}, arXiv
  e-prints, arXiv:2207.03532.
\newblock \doarXiv{2207.03532}

\bibitem[{{Welch} {et~al.}(2022{\natexlab{b}}){Welch}, {Coe}, {Diego},
  {Zitrin}, {Zackrisson}, {Dimauro}, {Jim{\'e}nez-Teja}, {Kelly}, {Mahler},
  {Oguri}, {Timmes}, {Windhorst}, {Florian}, {de Mink}, {Avila}, {Anderson},
  {Bradley}, {Sharon}, {Vikaeus}, {McCandliss}, {Brada{\v{c}}}, {Rigby},
  {Frye}, {Toft}, {Strait}, {Trenti}, {Sharma}, {Andrade-Santos}, \&
  {Broadhurst}}]{Welch22_nature}
{Welch}, B., {Coe}, D., {Diego}, J.~M., {et~al.} 2022{\natexlab{b}}, \nat, 603,
  815, \dodoi{10.1038/s41586-022-04449-y}

\bibitem[{{Welch} {et~al.}(2022{\natexlab{c}}){Welch}, {Coe}, {Zackrisson}, {de
  Mink}, {Ravindranath}, {Anderson}, {Brammer}, {Bradley}, {Yoon}, {Kelly},
  {Diego}, {Windhorst}, {Zitrin}, {Dimauro}, {Jimenez-Teja}, {Abdurro'uf},
  {Nonino}, {Acebron}, {Andrade-Santos}, {Avila}, {Bayliss}, {Benitez},
  {Broadhurst}, {Bhatawdekar}, {Bradac}, {Caminha}, {Chen}, {Eldridge},
  {Farag}, {Florian}, {Frye}, {Fujimoto}, {Gomez}, {Henry}, {Y. -Y Hsiao},
  {Hutchison}, {James}, {Joyce}, {Jung}, {Khullar}, {Larson}, {Mahler},
  {Mandelker}, {McCandliss}, {Morishita}, {Newshore}, {Norman}, {O'Connor},
  {Oesch}, {Oguri}, {Ouichi}, {Postman}, {Rigby}, {Ryan}, {Sharma}, {Sharon},
  {Strait}, {Strolger}, {Timmes}, {Toft}, {Trenti}, {Vanzella}, \&
  {Vikaeus}}]{Welch22_earendel}
{Welch}, B., {Coe}, D., {Zackrisson}, E., {et~al.} 2022{\natexlab{c}}, arXiv
  e-prints, arXiv:2208.09007.
\newblock \doarXiv{2208.09007}

\bibitem[{{Wen} \& {Han}(2015)}]{Wen2015}
{Wen}, Z.~L., \& {Han}, J.~L. 2015, \apj, 807, 178,
  \dodoi{10.1088/0004-637X/807/2/178}

\bibitem[{{Wen} {et~al.}(2012){Wen}, {Han}, \& {Liu}}]{Wen2012}
{Wen}, Z.~L., {Han}, J.~L., \& {Liu}, F.~S. 2012, \apjs, 199, 34,
  \dodoi{10.1088/0067-0049/199/2/34}

\bibitem[{{Wise} {et~al.}(2014){Wise}, {Demchenko}, {Halicek}, {Norman},
  {Turk}, {Abel}, \& {Smith}}]{wise2014}
{Wise}, J.~H., {Demchenko}, V.~G., {Halicek}, M.~T., {et~al.} 2014, \mnras,
  442, 2560, \dodoi{10.1093/mnras/stu979}

\bibitem[{{Wofford} {et~al.}(2016){Wofford}, {Charlot}, {Bruzual}, {Eldridge},
  {Calzetti}, {Adamo}, {Cignoni}, {de Mink}, {Gouliermis}, {Grasha}, {Grebel},
  {Lee}, {{\"O}stlin}, {Smith}, {Ubeda}, \& {Zackrisson}}]{Wofford16}
{Wofford}, A., {Charlot}, S., {Bruzual}, G., {et~al.} 2016, \mnras, 457, 4296,
  \dodoi{10.1093/mnras/stw150}

\bibitem[{{Xiao} {et~al.}(2019){Xiao}, {Galbany}, {Eldridge}, \&
  {Stanway}}]{Lin19}
{Xiao}, L., {Galbany}, L., {Eldridge}, J.~J., \& {Stanway}, E.~R. 2019, \mnras,
  482, 384, \dodoi{10.1093/mnras/sty2557}

\bibitem[{{Xiao} {et~al.}(2018){Xiao}, {Stanway}, \& {Eldridge}}]{Lin18}
{Xiao}, L., {Stanway}, E.~R., \& {Eldridge}, J.~J. 2018, \mnras, 477, 904,
  \dodoi{10.1093/mnras/sty646}

\bibitem[{{Zackrisson} {et~al.}(2011){Zackrisson}, {Rydberg}, {Schaerer},
  {{\"O}stlin}, \& {Tuli}}]{Zackrisson2011}
{Zackrisson}, E., {Rydberg}, C.-E., {Schaerer}, D., {{\"O}stlin}, G., \&
  {Tuli}, M. 2011, \apj, 740, 13, \dodoi{10.1088/0004-637X/740/1/13}

\end{thebibliography}
\bibliographystyle{aasjournal}

\end{document}